\documentclass[acmsmall,screen,review=false]{acmart}

\usepackage{graphicx}
\usepackage{amsmath}
\usepackage{colortbl}
\usepackage{arydshln}
\usepackage{multirow}
\usepackage{enumerate}
\usepackage{makecell}
\usepackage{amsmath,amsfonts}
\usepackage{algorithm}
\usepackage{algorithmic}
\usepackage{graphicx}
\usepackage{textcomp}
\usepackage{xcolor}
\usepackage{threeparttable}
\usepackage{caption}
\usepackage{hyperref}
\usepackage{varioref}
\usepackage{cleveref}
\usepackage{booktabs}
\usepackage[most]{tcolorbox}
\definecolor{lightgray}{rgb}{0.976, 0.976, 0.976}
\definecolor{ForestGreen}{rgb}{0.13, 0.55, 0.13}
\usepackage{adjustbox}

\newtcolorbox{mybox}{
  breakable,
  arc=3mm,
  colback=lightgray,
  colframe=black,
  boxrule=0.3mm,
}

\AtBeginDocument{
  }

\setcopyright{acmcopyright}
\copyrightyear{2023}
\acmYear{2023}
\acmDOI{XXXXXXX.XXXXXXX}
\acmJournal{TOSEM}

\begin{document}

\title{Self-collaboration Code Generation via ChatGPT}

\author{Yihong Dong}\authornote{Equal Contribution}
\email{dongyh@stu.pku.edu.cn}
\author{Xue Jiang}\authornotemark[1]
\email{jiangxue@stu.pku.edu.cn}
\author{Zhi Jin}
\email{zhijin@pku.edu.cn}
\author{Ge Li}
\email{lige@pku.edu.cn}

\authornote{Corresponding author}

\affiliation{%
  \institution{Key Laboratory of High Confidence Software Technologies (Peking University), Ministry of Education; School of Computer Science, Peking University, Beijing}
  \country{China}
}

\renewcommand{\shortauthors}{Dong et al.}

\begin{abstract}
Although Large Language Models (LLMs) have demonstrated remarkable code-generation ability, they still struggle with complex tasks. In real-world software development, humans usually tackle complex tasks through collaborative teamwork, a strategy that significantly controls development complexity and enhances software quality. Inspired by this, we present a self-collaboration framework for code generation employing LLMs, exemplified by ChatGPT. Specifically, through role instructions, 1) Multiple LLM agents act as distinct `experts', each responsible for a specific subtask within a complex task; 2) Specify the way to collaborate and interact, so that different roles form a virtual team to facilitate each other's work, ultimately the virtual team addresses code generation tasks collaboratively without the need for human intervention. To effectively organize and manage this virtual team, we incorporate software-development methodology into the framework. Thus, we assemble an elementary team consisting of three LLM roles (i.e., analyst, coder, and tester) responsible for software development's analysis, coding, and testing stages. We conduct comprehensive experiments on various code-generation benchmarks. Experimental results indicate that self-collaboration code generation relatively improves 29.9\%-47.1\% Pass@1 compared to the base LLM agent. Moreover, we showcase that self-collaboration could potentially enable LLMs to efficiently handle complex repository-level tasks that are not readily solved by the single LLM agent. \footnotetext{\url{https://github.com/YihongDong/Self-collaboration-Code-Generation}}
\end{abstract}



\maketitle

\section{Introduction}
Code generation aims to generate code that satisfies human requirements expressed in the form of some specification. Successful code generation can improve the efficiency and quality of software development, even causing changes in social production modes. Therefore, code generation has been a significant research hotspot in the fields of artificial intelligence, natural language processing, and software engineering. Recently, code generation has made substantial advancements in both academic and industrial domains \cite{codex, Subtoken-TranX, alphacode, CODEP}. In particular, LLMs have achieved excellent performance and demonstrate promising potential on code generation tasks \cite{nijkamp2022codegen, incoder, Zheng2023CodeGeeXAP, CDD}. 

Nonetheless, generating correct code for complex requirements poses a substantial challenge, even for experienced human programmers. Intuitively, humans, as social beings, tend to rely on collaborative teamwork when encountering complex tasks.
Teamwork through division of labor, interaction, and collaboration to solve complex problems, has been theorized to play an important role in dealing with complexity, as posited in both teamwork theory \cite{belbin2012team,katzenbach2015wisdom} and software engineering practice \cite{beck2001manifesto,mcchesney2004communication,demarco2013peopleware}. The benefits of collaborative teamwork are manifold: 1) It breaks down complex tasks into smaller subtasks, making the entire code generation process more efficient and controllable. 2) It assists with error detection and quality control. Team members can review and test the generated code, providing feedback and suggestions for improvement, thus reducing potential errors and defects. 
3) It ensures that the generated code is consistent with the expected requirements. Team members can offer different viewpoints to solve problems and reduce misunderstandings.

A straightforward way to implement collaborative teamwork entails training different models to handle the corresponding subtasks, subsequently conducting joint training to foster mutual understanding of behaviors to assemble them into a team \cite{PEER}. However, this training approach is costly, especially for LLMs. The scarcity of relevant training data further exacerbates the difficulty of achieving collaborative code generation.
Revolutionary advancements in artificial general intelligence (AGI), especially LLMs represented by ChatGPT \cite{ChatGPT}, provide a turning point. These LLMs perform commendably across tasks in various stages of software development, laying the groundwork for division of labor. Furthermore, LLMs use language as the foundation for input and output and align with human needs through instructions or prompts, offering the potential for inter-model interaction and collaboration. 

\begin{figure*}[h!]
\centering
\includegraphics[width=0.98\textwidth]{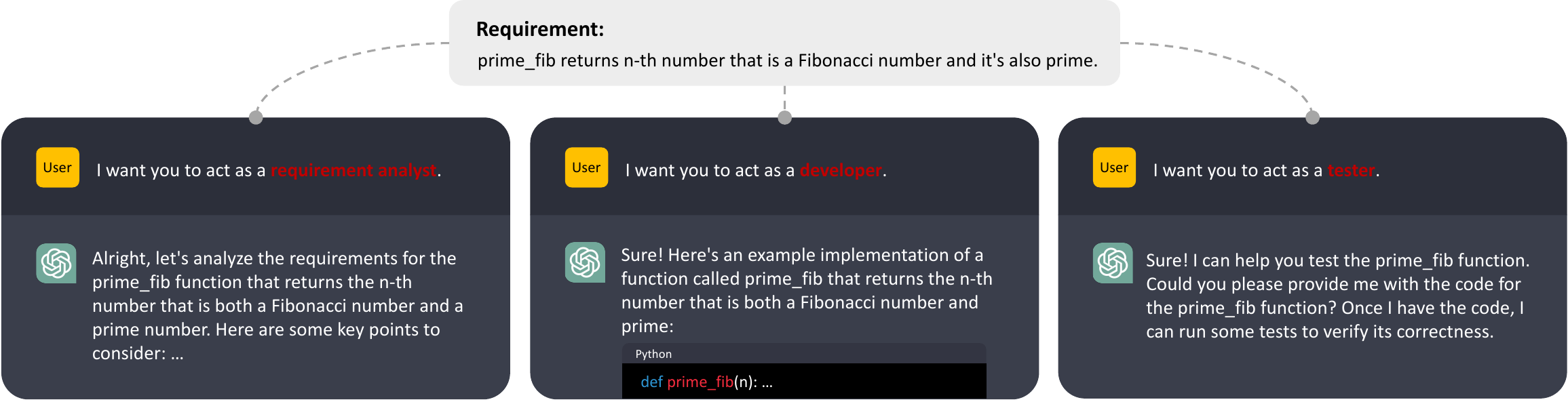}
\caption{An example of role-playing. 
Through role-playing, LLM transforms into an expert within a specific domain, delivering a professional-perspective response to the same requirement.
}
\label{role-playing}
\end{figure*}

To this end, we propose a self-collaboration framework aimed at guiding LLMs to collaborate with themselves, thereby dealing with complex requirements and boosting the performance of code generation. This framework is divided into two parts: division of labor and collaboration, both of which are dominated by role instructions.
First, role instructions achieve division of labor by allocating specific roles and responsibilities to LLMs. This strategy enables LLMs to think about and tackle tasks from the standpoint of the roles they play, transforming the original LLMs into domain `experts', as shown in Fig. \ref{role-playing}.
Second, role instructions control the interaction between roles, allowing otherwise isolated roles to form a virtual team and facilitate each other's work.

To promote efficient collaboration, we introduce software-development methodology (SDM) into self-collaboration framework. SDM provides a set of clearly defined stages, principles, and practices that serves to organize and manage teams effectively, ultimately controlling development complexity and improving software quality \cite{abrahamsson2002agile, lifecycle_models}. Following SDM, we instantiate an elementary team composed of three roles (i.e., analyst, coder, and tester) to achieve this goal. These roles adhere to an SDM-defined workflow where the stages of analysis, coding, and testing are performed sequentially, with each stage providing feedback to its predecessor. Specifically, the analyst breaks down requirements and develops high-level plans for guiding the coder; the coder creates or improves code based on the plans or the tester's feedback; the tester compiles test reports based on the coder's outcome and documents any issues found during testing. We employ three ChatGPT\footnote{The ChatGPT referenced throughout our paper defaults to the GPT-3.5 version.} agents to respectively play the three roles through role instructions, and then they collaborate to address code generation tasks under the guidance of self-collaboration framework. The primary contributions of our work can be summarized as follows: 

\begin{figure}[t!]
\centering
\includegraphics[width=0.9\textwidth]{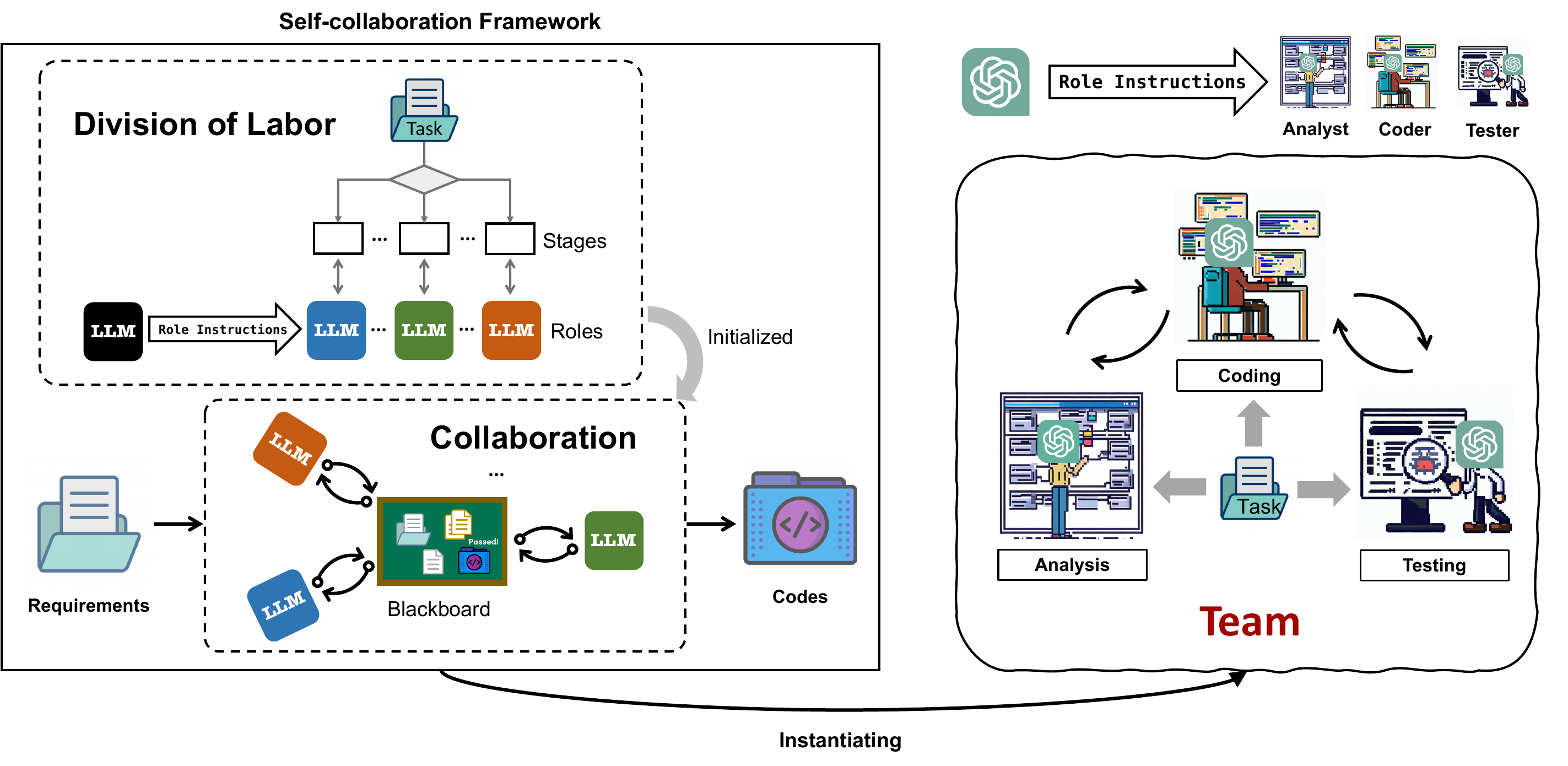}
\caption{Self-collaboration framework for code generation and its instance.}
\label{self-collaboration}
\end{figure}

\begin{enumerate}
    \item We propose a self-collaboration framework with role instruction, which allows LLM agents to collaborate with each other to generate code for complex requirements.
    \item Following software-development methodology, we instantiate an elementary team, which comprises three LLM roles (i.e., analyst, coder, and tester) responsible for their respective stages in the software development process.
    \item  Building on self-collaboration framework, the virtual team formed by ChatGPT (GPT-3.5) can achieve significant improvements compared to the single LLM agent on multiple code-generation benchmarks. 
    \item In some real-world scenarios, self-collaboration code generation exhibits notable effectiveness on more complex code generation tasks (such as repository-level code generation) that are challenging for the single LLM agent. 
\end{enumerate}

\section{Self-collaboration Framework}
Our self-collaboration framework consists of two parts: division of labor (DOL) and collaboration, which is shown in Fig. \ref{self-collaboration} (left). Given a requirement $x$, we propose to perform self-collaboration with LLMs to generate the output $y$. The task is defined as $\mathcal{T}: x \rightarrow y$.

\subsection{Division of Labor}
In DOL part, we leverage prior knowledge to decompose the process of solving a complex task into a series of stages $\mathcal{T} \Rightarrow \{\mathcal{S}_i \}^l_{i=1}$ and construct some distinct roles $\{R_j\}^m_{j=1}$ based on $\{\mathcal{S}_i \}^l_{i=1}$. Each stage $\mathcal{S}_i$ can be processed by one or more roles $R_j's$.\footnote{In our self-collaboration framework, each stage is managed by a specific type of roles, but the number of roles within that type can vary. The order of the stages determines the sequence in which different types of roles are executed, but the roles within a single stage operate in parallel.}

It is widely acknowledged that LLMs are sensitive to context, as they are trained to predict subsequent tokens based on preceding ones. Consequently, it is prevalent to control LLM generation using instructions or prompts \cite{instruction, FlanPaLM, GPT-4}. In order to achieve division of labor, we craft a specific type of instruction to assign roles and responsibilities to LLMs, which we refer to as role instructions. Specifically, we ask an LLM to act as a particular role that has a strong correlation with its responsibilities. Furthermore, we need to convey the detailed tasks, i.e. responsibilities,  this role should perform. 
In general, a clear and unverbose task description in the instruction could lead to LLM's behavior being more in line with your expectations. One case where it may not be necessary to outline a role's responsibilities is when the division of labor is common enough that matching roles can be found in reality.

Through role-playing, we can effectively situate LLM within a specific domain, thereby eliciting its expertise within that domain. Our empirical evidence suggests that this role-playing approach yields superior results compared to directly engaging the LLM in the task without a pre-defined contextual setting. Thus, role-playing can be harnessed as an efficient tool to enhance the performance of the LLM in specialized tasks.

Note that the role instruction only needs to be provided once at the initialization of each LLM agent to provide specific guidance on its behavior throughout subsequent interaction, thus enhancing the overall efficiency and clarity in collaboration.

\subsection{Collaboration}
After assigning roles to LLMs in DOL part, roles interact their outputs with other roles as the stages progress, refining the work and ensuring an accurate and thoughtful output $y$.
In collaboration part, we focus on facilitating effective interactions among distinct roles to ensure that they mutually enhance each other's work. 

The interaction among roles occurs in the form of natural language (NL), which is supported by the foundational aspects of the language model. We specify the role, information, and format that each role interacts with in role instructions, which allows the whole process of collaboration to be well controlled. The collaboration part can be formalized as follows:
\begin{equation}
    \underset{s_t}{\arg \max} \ \ P(s_{t}|s_{\{<t\}}, R_{m(S_t)}, x), \label{prob}
\end{equation}
where $s_{t}$ is the output of stage $\mathcal{S}_t$, $s_{\{<t\}}$ indicates the prerequisite-stage outputs of $\mathcal{S}_t$\footnote{Note that our self-collaboration framework can be parallelized if the relationship between stages $\{\mathcal{S}_i\}^l_{i=1}$ is not a straightforward linear relationship. In other words, if one stage does not depend on the results of another stage, then they can be executed in parallel. A specific scenario is that in a software development project including front-end, back-end, and database development, the analysis stage only needs to define the interface in advance and the corresponding coding stages can be carried out in parallel.}, and $R_{m(S_t)}$ represents the role corresponding to $\mathcal{S}_t$. We consider the computation of $P(s_{t}|s_{<t}, R_{m(S_t)}, x)$ as the collaboration, wherein role $R_{m(S_t)}$ collaborates with the roles of each preceding stage to generate $s_{t}$. Output $y$ is iteratively updated along with the progression of $\mathcal{S}_t$:
\begin{equation}
    y_t = f(s_t, y_{<t}), \label{update}
\end{equation}
where $f$ is an update function. Once the end condition is satisfied\footnote{The end condition is defined by prior knowledge, and an example can be found in the last paragraph of Section \ref{Instance}.}, the final output $y$ is derived. To coordinate collaboration between different roles, we set up a shared blackboard \cite{Blackboard}, from which each role exchanges the required information to accomplish their respective tasks $s_t$ via Eq. \eqref{prob}. The pseudocode of our self-collaboration framework is outlined in Algorithm \ref{algorithm1}.

\begin{algorithm}[t!]
\caption{Pseudocode of self-collaboration framework.}\label{algorithm1}
\begin{algorithmic}[1]
\REQUIRE{Requirement $x$, Task $\mathcal{T}$, and LLM $\mathcal{M}$.}
\ENSURE{Output $y$.}\\
    \textcolor{gray}{\textit{\# \textbf{DOL Part}}}
\STATE Initial Stages $\{\mathcal{S}_i \}^l_{i=1}$ according to task $\mathcal{T}$.
\STATE Initial Roles $\{R_j\}^m_{j=1}$ for LLMs $\mathcal{M}'s$ based on stages $\{\mathcal{S}_i \}^l_{i=1}$ and role instructions.\\
    \textcolor{gray}{\textit{\# \textbf{Collaboration Part}}}
    \STATE Initial blackboard $\mathcal{B}$ and index $t$.
\REPEAT
    \STATE Obtain $s_{\{<t\}}$ from $\mathcal{B}$.
\STATE Sample $s_t$ via Eq. \eqref{prob}.
    \STATE Add $s_t$ to $\mathcal{B}$.
    \STATE Compute $y_t$ via Eq. \eqref{update}.
    \STATE Update $y$ and $t$.
\UNTIL{End condition is satisfied}
\RETURN{$y$}
\end{algorithmic}
\end{algorithm}

\section{Instance}
\label{Instance}
We introduce the classic waterfall model \cite{Waterfall} from software-development methodology into self-collaboration framework to make the teamwork for code generation more efficient. Specifically, we design a simplified waterfall model consisting of three stages, i.e. analysis, coding, and testing, as an instance for self-collaboration code generation.
The workflow of this instance follows the waterfall model flowing from one stage to the next, and if issues are found, it returns to the previous stage to refine. Thus, we establish an elementary team, comprising an analyst, coder, and tester, responsible for the analysis, coding, and testing stages, as illustrated in Fig. \ref{self-collaboration} (right). These three roles are assigned the following tasks:

\textbf{Analyst.} The goal of the analyst is to reduce the difficulty of coding by abstracting and decomposing the task from a high level, rather than delving into the details of the implementation. Given a requirement $x$, the analyst breaks $x$ down into several easily solvable subtasks to facilitate the division of functional units and develops a high-level plan to guide the coder in writing the code.

\textbf{Coder.} As the central role of this team, the coder is responsible for writing the code, but its work is carried out with the assistance and supervision of the analyst and tester. Thus, we assign two responsibilities to the coder: 1) Write code that fulfills the specified requirements, adhering to the plan provided by the analyst.
2) Repair or refine code, taking into account the feedback of test reports feedbacked by the tester.

\begin{figure}[th!]
\centering
\includegraphics[width=0.95\textwidth]{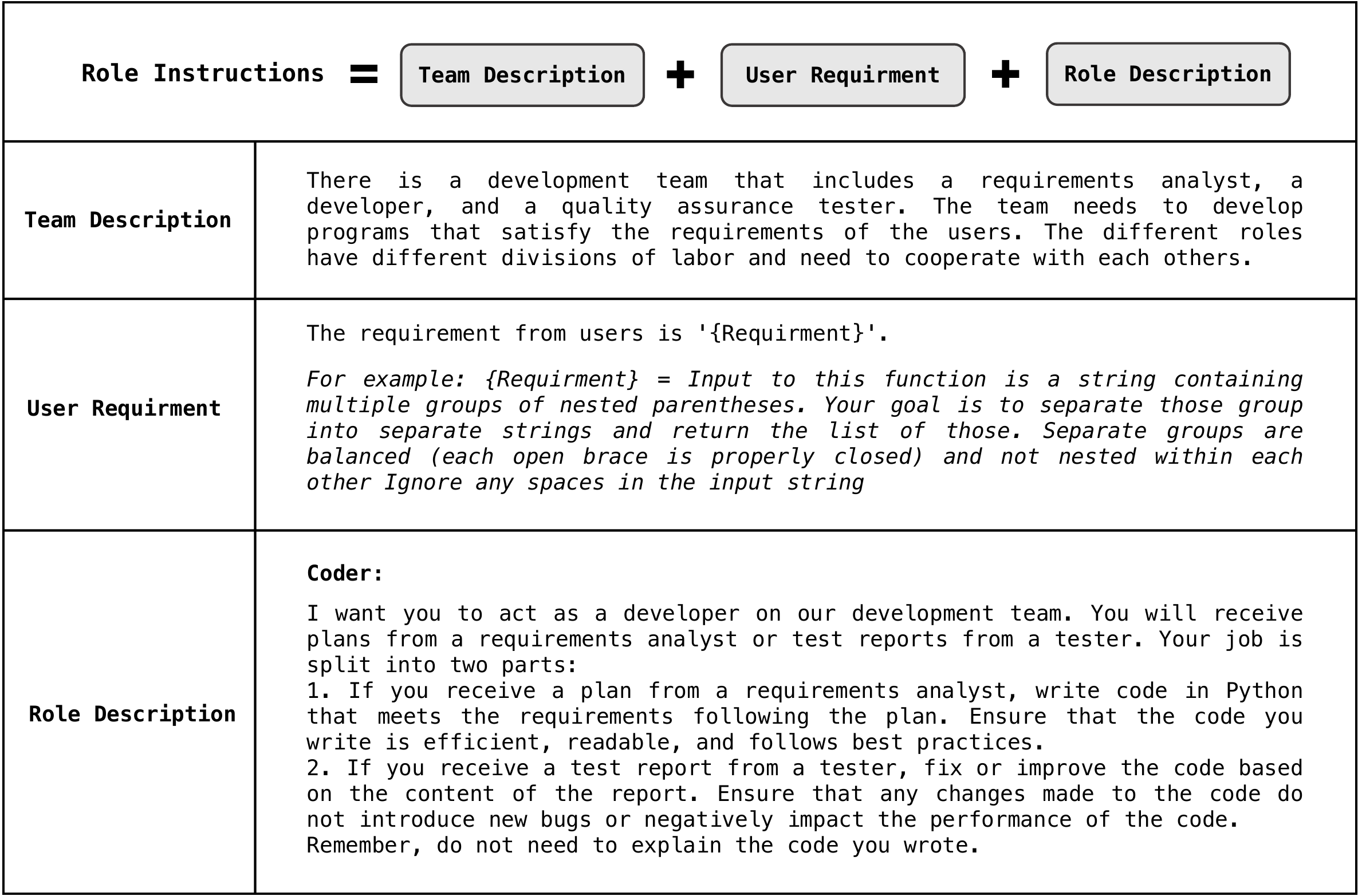}
\caption{An example of role instruction for coder in the instance of self-collaboration framework.}
\label{role_instructions}
\end{figure}

\textbf{Tester.}
The tester is responsible for inspecting the code and generating a test report on various aspects of functionality, readability, and maintainability to help the coder improve the quality of its code. 
Rather than directly introducing a compiler and test cases to execute the code, we use the model to simulate the testing process and produce test reports, thereby avoiding external efforts.

We customize role instructions for LLMs (exemplified by ChatGPT) to play the three roles. The role instruction includes not only the role description (role and its responsibilities) but also the team description and the user requirements, which will work together to initialize the ChatGPT agent, thereby setting the behavior of ChatGPT. An example of role instruction for coder is shown in Fig. \ref{role_instructions}. In addition, interactions occur between roles responsible for two successive stages, and we limit the maximum interaction to $n$. 
We update the output $y_t$ only when the stage $S_t$ is coding, and this workflow terminates upon $n$ is reached or the tester confirms that no issues persist with $y_t$. When dealing with $y_t$ composed of multiple components, it is recommended to ask LLMs to generate outputs in JSON format directly, which can reduce omissions and enhance the quality of generation outputs.

\section{Evaluation}
We aim at answering the following research questions (RQs):
\begin{itemize}
    \item \textbf{RQ1}: What is the performance of self-collaboration approach compared to the various baselines on public code-generation benchmarks?
    \item \textbf{RQ2}: What is the effect of roles in self-collaboration? Specifically, it can be divided into three questions: 1. What is the contribution of each role in the virtual team? 2. What is the effect of other virtual teams? 3. What is the effect of role-playing? 
    \item \textbf{RQ3}: What is the performance of self-collaboration based on different LLMs, especially the most powerful LLM GPT-4?
    \item \textbf{RQ4}: What is the impact of interaction numbers for self-collaboration?
    \item \textbf{RQ5}: What are the results of more detailed analysis (specifically, error analysis and cost analysis) for self-collaboration? 
    \item \textbf{RQ6}: How does self-collaboration work in repository-level software development scenarios and how does it perform?
\end{itemize}

\subsection{Experiment Setup}

\subsubsection{Benchmarks} 
We perform a comprehensive evaluation on six code-generation benchmarks to demonstrate the efficacy of self-collaboration. In addition, we also collect several projects in real-world development environments to showcase the excellent performance of self-collaboration on more complex code generation tasks.
\begin{itemize}
    \item \textbf{MBPP} (sanitized version) \cite{mbpp} contains 427 manually verified Python programming tasks, covering programming fundamentals, standard library functionality, and more. Each task consists of an NL description, a code solution, and 3 automated test cases.
    \item \textbf{HumanEval} \cite{codex} consists of 164 handwritten programming tasks, proposed by OpenAI. Each task includes a function signature, NL description, use cases, function body, and several unit tests (average 7.7 per task).
    \item \textbf{MBPP-ET} and \textbf{HumanEval-ET} \cite{CodeScore} are expanded versions of MBPP and HumanEval with over 100 additional test cases per task. This updated version includes edge test cases that enhance the reliability of code evaluation compared to the original benchmark.  
    \item \textbf{APPS} \cite{APPS} consists of problems collected from different open-access coding websites such as Codeforces, Kattis, and more. Problems are more complicated, as the average length of a problem is 293.2 words (about 5-25 times higher than preceding benchmarks). Each problem has multiple specifically designed test cases, and the average number of test cases is 21.2. 
    \item \textbf{CoderEval} \cite{CoderEval} contains four types of code in addition to the standalone code, i.e., Plib-depend, Class-depend, File-epend, and Project-depend, to simulate the dependencies in real-world development.
\end{itemize}

\subsubsection{Baselines} 
In this paper, we compare self-collaboration with three kinds of baselines: LLMs customized for code, generalist LLMs, and prompting approaches.

\paragraph{\textbf{I. LLMs customized for code} represent the LLMs specifically designed to solve code generation tasks and optimized on a large amount of code data, which helps to locate the performance level of our approach. This kind of baseline includes:}
\begin{itemize}
    \item \textbf{AlphaCode (1.1B)} \cite{alphacode} is a transformer-based code generation model, which is trained on selected public codes before July 14, 2021, and can solve some basic code generation problems.
    \item \textbf{Incoder (6.7B)} \cite{incoder} is a unified generation model that allows left-to-right code generation and code infilling/editing by the causal mask language modeling training objective. 
    \item \textbf{CodeGeeX (13B)} \cite{Zheng2023CodeGeeXAP} is a large-scale multilingual code generation model with 13 billion parameters, pre-trained on a large code corpus of more than 20 programming languages.
    \item \textbf{StarCoder (15.5B)} \cite{starcoder}  is a Code LLM trained on permissively licensed data from GitHub, including from 80+ programming languages, Git commits, GitHub issues, and Jupyter notebooks.
    \item \textbf{CodeGen (16.1B)} \cite{nijkamp2022codegen} is a LLM trained on NL and programming data for conversation-based program synthesis. In this paper, we employ the CodeGen-Multi version.
    \item \textbf{PaLM Coder (540B)} \citep{PaLM} is finetuned from PaLM 540B on code, where PaLM uses an ML system, named Pathways, that enables highly efficient training of LLMs across thousands of accelerator chips.
    \item \textbf{CodeX (175B)} \citep{codex}, also known as code-davinci-002, is fine-tuned from davinci 175B \citep{GPT3} on multilingual code data with code-completion tasks. CodeX is also the backbone model that powers Copilot \citep{Copilot} (a well-known commercial application).
    \item \textbf{CodeX (175B) + CodeT} \cite{codet} is a previous state-of-the-art (SOTA) approach before GPT-4. CodeT employs LLMs to automatically generate test cases for code samples. It executes code samples with these test cases and conducts a dual execution agreement, taking into account output consistency against test cases and output agreement among code samples.
    \item \textbf{CodeLlama (34B)} \cite{codellama}
    is an open foundational model for code generation tasks, derived from continuous training and fine-tuning based on Llama 2 \cite{Llama2}.
\end{itemize}

\paragraph{\textbf{II. Generalist LLMs} represent the LLMs that are trained on data widely collected from Internet and show strong performance on a variety of tasks, which are used as the base model of our approach. This kind of baseline includes:}
\begin{itemize}
    \item \textbf{ChatGPT} \citep{ChatGPT} is a sibling model to InstructGPT \citep{InstructGPT}, which is trained to follow an instruction in a prompt and provide a detailed response. We access ChatGPT through OpenAI's API. Since ChatGPT receives regular updates, we employ a fixed version `gpt-3.5-turbo-0301' as our base model, which will not receive updates, to minimize the risk of unexpected model changes affecting the results. 
    \item \textbf{GPT-4} \citep{GPT-4} is a large-scale, multimodal model which can accept image and text inputs and produce text outputs. GPT-4 exhibits human-level performance on various benchmarks.
\end{itemize}

\paragraph{\textbf{III. Prompting approaches} here represent the general prompting approaches as well as those specifically designed for code generation, representing a category of our related work. This kind of baseline includes:}
\begin{itemize}
    \item  \textbf{Chain-of-thought (CoT)} \cite{cot} generates a chain of thought for each question and then generates the corresponding code. For CoT, we use the instruction ``Let's think step by step.'' \cite{KojimaGRMI22} to implement it.
    \item \textbf{Self-planning} \cite{Self-planning} and \textbf{Self-debugging} \cite{debug} teach LLMs to perform planning and debugging with few-shot prompting. We use the prompts provided in their original papers to implement them. 
    \item \textbf{Iter-improving} is a baseline approach proposed in this paper, which considers allowing the base model to continuously improve the generated code until it can no longer be modified. This approach is used to demonstrate that the effectiveness of our method is not solely due to multiple improvements to the output of LLMs. We use the instruction `Please improve this code' to implement iter-improving, and the maximum number of iterations is set to 10.    
\end{itemize}

\subsubsection{Evaluation Metric.} 
In this paper, we mainly focus on \textbf{Pass@1}: the probability that the model solves the problem in one attempt, since in real-world scenarios we usually only consider one generated code.
We adopt the unbiased variant of Pass@1 \citep{codex} to measure the functional correctness of top-1 generated codes by executing test cases, which can be formulated as:
\begin{equation}
    \operatorname{Pass@1} = \mathop{\mathbb{E}}\limits_{\operatorname{Problems}}\begin{bmatrix}\chi(\text{generated code)}\end{bmatrix}.
\end{equation}
where $\mathop{\mathbb{E}}\limits_{\operatorname{Problems}}$ denotes the expectation of all problems and $\chi$ is an indicator function that outputs 1 if generated code passes all test cases of the corresponding problem, otherwise 0.

\subsubsection{Implementation Details.} 
For self-collaboration and all prompting approaches, we employ ChatGPT as the base model. We access ChatGPT using the `gpt-3.5-turbo' API with fixed version `0301', which will not receive updates. For self-collaboration, the maximum number of interactions between roles is limited to 4. 
In all experiments, we set max tokens to 512 and temperature to 0 for code generation. We only reference the results reported in their original papers for AlphaCode (1.1B) \cite{alphacode} and PaLM Coder (540) \cite{PaLM}, which are inaccessible to us. The results of all other baselines are evaluated under the same settings of self-collaboration for fairness. All these results are comparable with the results reported in their original paper.

\begin{table}[h!]
\caption{Comparison of self-collaboration and baselines, where the green highlights indicate the improvements in comparison to ChatGPT (GPT-3.5).}
\centering
\resizebox{\textwidth}{!}{
\begin{tabular}{@{}lcccc@{}}
\toprule
Approach  & \multicolumn{1}{c}{HumanEval} & \multicolumn{1}{c}{HumanEval-ET} & \multicolumn{1}{c}{MBPP} & \multicolumn{1}{c}{MBPP-ET} \\ \midrule
\textbf{LLMs Customized for Code}\\
AlphaCode (1.1B) \cite{alphacode} & 17.1 & - & - & - \\
Incoder (6.7B) \cite{incoder} & 15.2 & 11.6  & 17.6 & 14.3 \\
CodeGeeX (13B) \cite{Zheng2023CodeGeeXAP} & 18.9 & 15.2 & 26.9 & 20.4 \\
StarCoder (15.5B) \cite{starcoder} & 34.1 & 25.6 & 43.6 & 33.4 \\
CodeGen-Mono (16.1B) \cite{nijkamp2022codegen}  & 32.9 & 25.0 & 38.6 & 31.6 \\
PaLM Coder (540B) \cite{PaLM} & 36.0 & - & 47.0 & -\\
CodeLlama (34B) \cite{codellama} & 48.2 & 36.8 & 55.9 & 42.1 \\
CodeX (175B) \cite{codex} & 47.0 & 31.7 & 58.1 & 38.8 \\
CodeX (175B) + CodeT \cite{codet} & 65.8 & 51.7 & 67.7 & 45.1
\\ 
\hdashline
\textbf{Generalist LLMs} \\
ChatGPT (GPT-3.5) \cite{ChatGPT} & 57.3 & 42.7 & 52.2 & 36.8 \\
GPT-4 \cite{GPT-4} & 67.6 & 50.6 & 68.3 & 52.2 \\
\hdashline
\textbf{Prompting Approaches} \\
Zero-shot CoT \cite{cot} & 44.6 & 37.2 & 46.1 & 34.8 \\
Self-planning \cite{Self-planning} & 65.2 & 48.8 & 58.6 & 41.5 \\
Self-debugging \cite{debug} & 61.6 & 45.8 & 60.1 & 52.3 \\
Iter-improving & 55.5 & 46.3 & 45.7 & 32.8 \\
\hdashline
Ours & \textbf{74.4} \ \ (\textcolor{ForestGreen}{ $\uparrow$ 29.9\%}) & \textbf{56.1} \ \ (\textcolor{ForestGreen}{ $\uparrow$ 31.4\%}) & \textbf{68.2} \ \ (\textcolor{ForestGreen}{ $\uparrow$ 30.7\%}) & \textbf{49.5} \ \ (\textcolor{ForestGreen}{ $\uparrow$ 34.6\%}) \\ 
\bottomrule
\end{tabular}}
\label{main_result}
\end{table}

\subsection{RQ1: Self-collaboration vs. Baselines} 
We compare self-collaboration with baselines in the `NL + signature + use cases' setting\footnote{This setting provides requirements, function signature, and use cases as input to generate code. which is commonly used for baselines, so we also adopt this setting for comparison. However, using only requirements (NL-only) as input is more consistent with real-world development scenarios and more challenging. Therefore, in other experiments, we use the `NL-only' setting if not specified. Detailed discussions can be found in Appendix \ref{detailed Settings and Baselines}.}, as shown in Table \ref{main_result}.
Experimental results show that self-collaboration code generation based on ChatGPT (GPT-3.5), with a three-person team (including analysts, coders, and testers), achieves SOTA performance across the four code generation benchmarks.
When compared to ChatGPT, the improvement offered by our framework is substantial, with a relative increase ranging from 29.9\% to 34.6\%. It is noteworthy that the self-collaboration code generation yields more significant improvements on the datasets associated with extended test cases, namely HumanEval-ET and MBPP-ET. This suggests that our self-collaboration framework can effectively assist base LLMs in generating reliable code. This enhancement may be attributed to the collaborative team's ability to consider a wider range of boundary conditions and address common bugs. 

Compared to zero-shot CoT, Iter-improving, and two concurrent works (self-planning and self-debugging), self-collaboration also performs significantly better than these prompting baselines. 
It is worth mentioning that CoT is a successful prompting technique that solves the reasoning problem by generating intermediate reasoning steps, but CoT is not suitable for code generation \cite{Self-planning,Text-to-SQL}. 
We find that the performance of Iter-improving is usually lower than the single LLM agent with ChatGPT except for HumanEval-ET. This is because ChatGPT tends to optimize for edge test cases. While this helps in HumanEval-ET, it sometimes leads to unnecessary modifications of correct edge cases. Such adjustments might be safer in some scenarios, like throwing exceptions, but deviate from expected output.

We evaluate self-collaboration on a more algorithmically intricate benchmark, namely, APPS. This benchmark comprises three levels of difficulty: introductory, interview, and competition. In this experiment, we select the first 1000 code generation tasks from the most challenging two levels: interview and competition. Following the settings in CodeT \cite{codet}, we use NL in APPS as a comment and then concatenate the signature `\textit{def solution(stdin : str) → str :}' as input to the model. Experimental results shown in Table \ref{extend_apps} indicate that self-collaboration enhances the performance of ChatGPT substantially and exceeds the performance of the previous SOTA approach CodeX (175B) + CodeT \cite{codet}.

\begin{table}[t!]
\caption{The performance of self-collaboration code generation on APPS, where the green highlights indicate the improvements in comparison to ChatGPT (GPT-3.5).}
\label{extend_apps}
\centering
\normalsize{
\begin{tabular}{lcc}
\toprule
Approach                     & Competition & Interview \\ \midrule
CodeX (175B) + CodeT \cite{codet}                        & 2.2         & 8.1       \\
ChatGPT (GPT-3.5) \cite{ChatGPT}  & 1.2         & 6.8       \\
\hdashline
Ours & \textbf{3.4}  \ \ (\textcolor{ForestGreen}{ $\uparrow$ 183.3\%})        & \textbf{10.7}  \ \ (\textcolor{ForestGreen}{ $\uparrow$ 57.4\%})        \\ \bottomrule
\end{tabular}
}
\end{table}

To evaluate self-collaboration on a more challenging and realistic benchmark, we conduct experiments on CoderEval  \cite{CoderEval} based on ChatGPT (GPT-3.5), following the settings in its paper. As shown in Table \ref{CoderEvalExperiment}, self-collaboration code generation substantially outperforms the single ChatGPT, achieving relative improvements of 47.1\% on Pass@1.

\begin{table}[h!]
\caption{The performance of self-collaboration code generation on CoderEval.}
\label{CoderEvalExperiment}
\resizebox{\textwidth}{!}{
\begin{tabular}{@{}lcccccc@{}}
\toprule
Approach          & \multicolumn{1}{c}{Standalone} & \multicolumn{1}{c}{Plib-depend} & \multicolumn{1}{c}{Class-depend} & \multicolumn{1}{c}{File-epend} & \multicolumn{1}{c}{Project-depend} & \multicolumn{1}{c}{Average} \\ \midrule
ChatGPT (GPT-3.5) & 35.87                          & 21.43                           & 8.73                             & 16.91                          & 9.57                               & 19.82                       \\
\hdashline
Ours & \textbf{47.63} & \textbf{38.1}  & \textbf{21.82} & \textbf{20.59}  & \textbf{13.04} & \textbf{29.14} \ \ (\textcolor{ForestGreen}{ $\uparrow$ 47.1\%})                      \\ \bottomrule
\end{tabular}}
\end{table}

\begin{table}[t!]
\setlength\tabcolsep{6pt}
\caption{Effectiveness of ChatGPT roles in self-collaboration code generation, where the green highlights indicate the improvements in comparison to Coder.}
\label{ablation}
\centering
\normalsize{
\begin{threeparttable}
\begin{tabular}{@{}lcc@{}}
\toprule
Roles & HumanEval & HumanEval-ET \\ \midrule
\textbf{Ablation of Roles} \\
Coder  & 45.1    & 35.3\\
+ Analyst          & 54.9 \ \ (\textcolor{ForestGreen}{ $\uparrow$ 21.8\%})     &  45.2 \ \ (\textcolor{ForestGreen}{ $\uparrow$ 28.1\%})  \\
+ Tester            & 56.8 \ \ (\textcolor{ForestGreen}{ $\uparrow$ 26.0\%})    & 45.2 \ \ (\textcolor{ForestGreen}{ $\uparrow$ 28.1\%}) \\
+ Analyst + Tester & \textbf{63.5} \ \ (\textcolor{ForestGreen}{ $\uparrow$ 40.8\%})   &  \textbf{51.9} \ \ (\textcolor{ForestGreen}{ $\uparrow$ 47.1\%}) \\
\hdashline
\textbf{Different Team Configurations} \\
Analyst + Coder + Tester + Compiler & 64.6 & 50.0\\
Driver + Observer (Pair Programming)  & 57.4  & 44.6\\ 
\hdashline
\textbf{The Role of Role-playing} \\
Few-shot Prompting (4-shot) & 58.1    & 47.5   \\
Instruction (zero-shot)  & 60.0     & 47.5\\  
Role Instruction (Ours)     & \textbf{64.4}     & \textbf{51.9}   \\ \bottomrule
\end{tabular}
 \begin{tablenotes}
        \item[I] \normalsize{We sample 4 HumanEval tasks to construct examples for few-shot prompting, which are excluded in evaluations of  `The Role of Role-playing'.}
 \end{tablenotes}
\end{threeparttable}}
\end{table}

\subsection{RQ2: The Effect of Roles in Self-collaboration} \label{role_playing_exp}
We investigated code generation relying solely on NL descriptions. 
In this setting, we explore the effect of roles in self-collaboration in three ways: 1) Conducting ablation studies with the three-roles team we assembled (analyst, coder, and tester). 2) Exploring alternative team configurations, i.e., a team incorporating a compiler, and a team of pair programming. The team with the compiler constructs the use cases in HumanEval as public test cases and obtains the execution results through compiler to help the tester generate reports. In the team with pair programming, we use a two-agent team of driver and observer for code generation. 3) Implementing the analysis, coding, and testing phases directly without role-playing to validate the role of role-playing. To implement these tasks, we used both instruction and few-shot prompting strategies. Instruction (zero-shot), which involves instructions with the role-playing component excised, and few-shot prompting, which provides examples demonstrating the tasks of analyzing, coding, and testing. Details of instructions and prompts are provided in Appendix~\ref{detailed Prompt or instruction}. The experimental results are shown in Table \ref{ablation}.

First, the results show that the performance substantially improves when compared to employing only the coder role after forming a team, regardless of whether it is a two-role or three-role team. The coder-analyst-tester team achieved the best results on HumanEval and HumanEval-ET benchmarks, with relative improvements of 40.8\% and 47.1\%, respectively. 

Second, we find that the performance of `Analyst + Coder + Tester + Compiler' is comparable to that of `Analyst + Coder + Tester'. The reason may be that some mistakes can already be resolved by the tester without relying on the complier's results. Moreover, the team of pair programming is superior to the single LLM agent, but still performed slightly worse compared to the team of `Analyst + Coder + Tester'.

Third, the results show that role-playing approach substantially outperforms the baselines without role-playing. We suppose that the possible reason for the better performance of role-playing LLMs is that it provides a specific context that constrains the generation space of LLMs, making it reason within the constraints of the scenario, generating responses that align with the perspectives that LLM in that role might have. Therefore, role-playing serves to evoke the latent abilities of LLMs instead of directly improving LLM's abilities. Moreover, in two scenarios without role-playing, we observe that instruction (zero-shot) is slightly better than few-shot prompting. We identified two potential factors that could lead to this observation. 1) Few-shot prompting may bias the LLMs' understanding of human intent due to the limited selection of examples that might not fully reflect intent. 2) The long prompt (about 14 times the instruction length in the experiment) used in few-shot prompting could hinder the LLMs' effective extraction of relevant information. 

\begin{figure}[h!]
\centering
\includegraphics[width=0.9\textwidth]{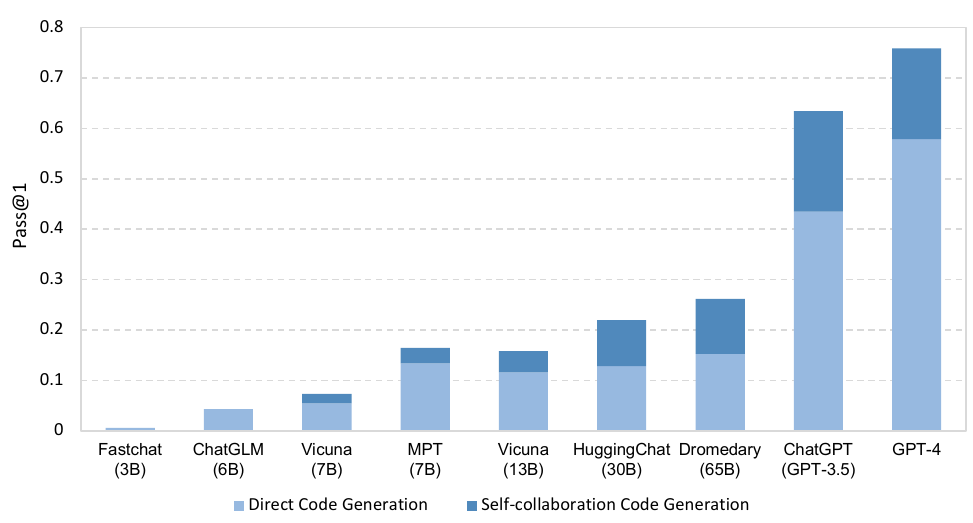}
\caption{Self-collaboration capacities of different LLMs.}
\label{different_model}
\end{figure}

\subsection{RQ3: Self-collaboration on Different LLMs}
In addition to ChatGPT, we apply self-collaboration to seven open-source LLMs, including Fastchat-3B \citep{Fastchat}, ChatGLM-6B \citep{ChatGLM}, MPT-7B \citep{MosaicML2023Introducing}, Vicuna-7B/13B \citep{Vicuna}, HuggingChat-30B \citep{HuggingChat}, and Dromedary-65B \citep{Dromedary}, and the most powerful LLM GPT-4 \cite{GPT-4}. Our objective is to evaluate the efficacy of self-collaboration in tackling complex tasks, specifically those challenging for the single LLM agent. For such tasks on HumanEval, we employ the self-collaboration strategy as the solution. As illustrated in Fig. \ref{different_model}, the coding ability of chat-based LLMs generally exhibits an increasing trend with the enlargement of model sizes. The self-collaboration capability starts to manifest itself around the 7B parameters and subsequently continues to escalate, serving to evoke latent intelligence within LLMs.

For the most powerful LLM GPT-4, we conduct additional experiments to evaluate self-collaboration code generation based on GPT-4, following the settings in the GPT-4 technical report \cite{GPT-4}. The experimental results are shown in Table \ref{extend_gpt4}, and we can find that the enhancement effect of self-collaboration on GPT-4 is significant.

\begin{table}[h!]
\centering
\caption{The performance of self-collaboration code generation with GPT-4. The result in brackets is reported on the GPT-4 technical report \cite{GPT-4}.}
\label{extend_gpt4}
\normalsize{
\begin{tabular}{lcccc}
\toprule
Approach                   & HumanEval  & HumanEval-ET & MBPP & MBPP-ET \\ \midrule
GPT-4 \cite{GPT-4}                     & 67.7 (67.0) & 50.6         & 68.1 & 49.2    \\
GPT-4 + Self-collaboration & 90.2       & 70.7         & 78.9 & 62.1    \\ \bottomrule
\end{tabular}
}
\end{table}

\begin{figure*}[h!]
\centering
\includegraphics[width=0.78\textwidth]{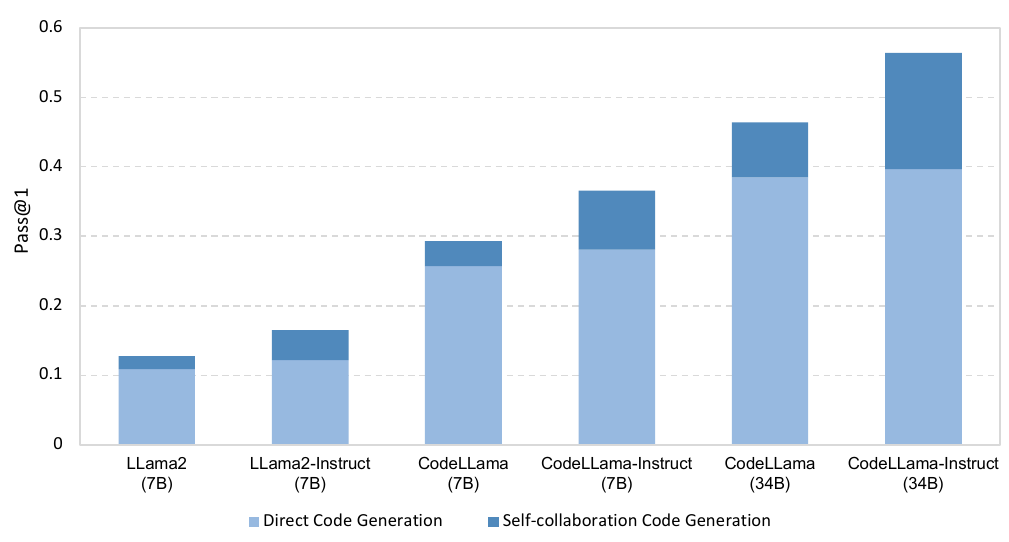}
\centering
\caption{Self-collaboration capacities of CodeLlama series.}
\label{CodeLlamaSeries}
\end{figure*}

To figure out the abilities required for self-collaboration, we conduct experiments on a series of models including CodeLlama 7B and 34B \cite{codellama}, Llama2 \cite{Llama2} (the base model of CodeLlama) and their Instruct version (model with instruction tuning). As illustrated in Figure \ref{CodeLlamaSeries}, the enhancement observed in Llama2-7B using self-collaboration falls short when compared to CodeLlama-7B. This discrepancy emphasizes the critical role of domain-specific expertise. The performance improvement of the Instruct version using self-collaboration generally exceeds that of the original version, highlighting the significance of in-context learning capabilities. Furthermore, the advancement in CodeLlama-34B using self-collaboration eclipses that of CodeLlama-7B on both two versions, underscoring the importance of reasoning abilities. Therefore, self-collaboration may require the following abilities for LLMs, including strong domain-specific expertise ability for role-playing, strong in-context learning ability to follow instructions, and strong reasoning ability to solve problems effectively.

\subsection{RQ4: The Effect of Interaction}
To evaluate the impact of interaction on self-collaboration code generation, we control the maximum number of interactions (MI) between roles in this experiment, with the results shown in Table \ref{MI}. The MI value equal to zero signifies a complete absence of interaction between the roles involved. This means that the code generated by a particular coder does not receive feedback from other roles, so the result is the same as employing a coder only. The MI value increases from 0 to 1 for the most significant performance improvement. This result shows that for these four benchmarks, our approach solves most of the tasks within two rounds (i.e., one round of interaction). The continued increase in the MI value beyond the initial round yields diminishing returns in terms of improvement; however, there is still a consistent enhancement observed. This suggests that more complex tasks are undergoing continuous improvement. In general, higher MI values yield better outcomes. Nonetheless, due to the maximum token constraint of ChatGPT, our exploration was limited to the MI value from 1 to 4 rounds.

\begin{table}[h!]
    \caption{The effect of maximum interaction (MI) for self-collaboration code generation.}
\label{MI}
\centering
\normalsize{
\begin{tabular}{@{}ccccc@{}}
\toprule
MI & HumanEval & HumanEval-ET & MBPP  & MBPP-ET \\ \midrule
0            & 45.1    & 35.3       & 47.5 & 34.3  \\
1            & 60.4     & 50.7        & 53.0 & 38.0   \\  2                  & 62.2     & 50.7        & 54.1 & 38.5   \\  4                  & 63.5     & 51.9        & 55.5 & 40.8   \\ \bottomrule
\end{tabular}}
\end{table}

\begin{figure*}[h!]
\centering
\includegraphics[width=\textwidth]{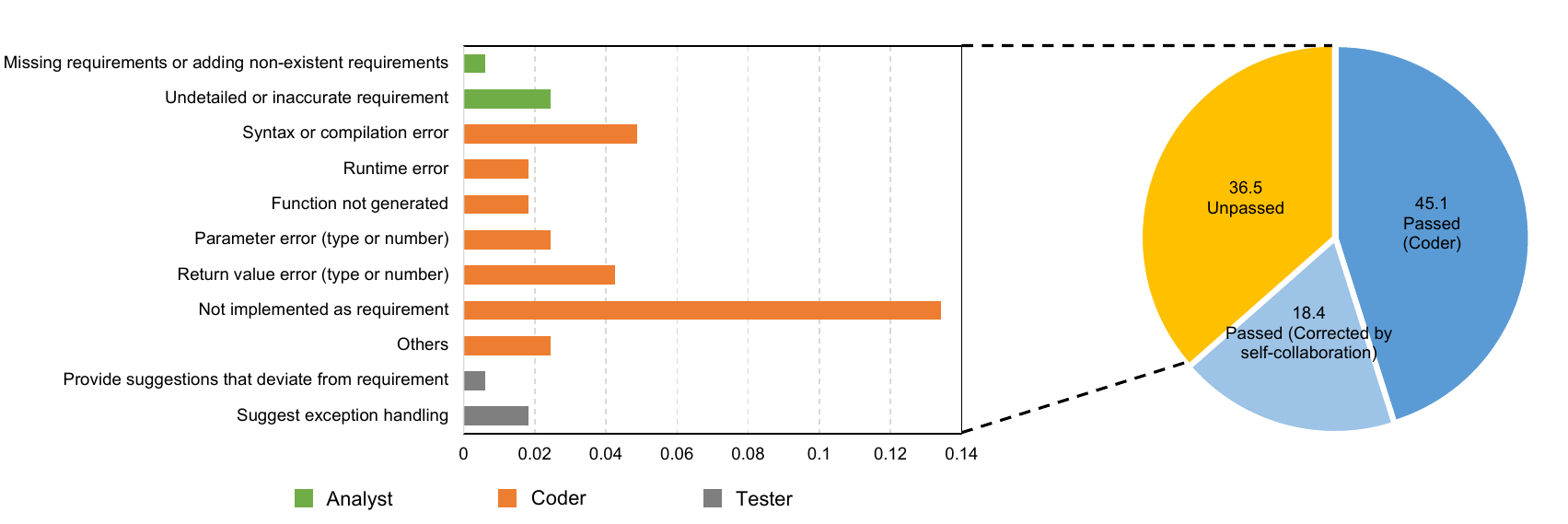}
\caption{Error Analysis for Self-collaboration.}
\label{error_analysis}
\end{figure*}

\subsection{RQ5: Analysis for Self-collaboration}
\label{E_Analysis}
In this section, we further analyze the errors generated by self-collaboration and the cost of self-collaboration compared to other prompting approaches.

\subsubsection{Error Analysis}
To deepen the understanding of error modes for self-collaboration, we conducted a qualitative error analysis. On HumanEval dataset, we classify tasks into three categories: the first category includes tasks that can be correctly generated by the coder; the second category includes tasks that cannot be correctly generated by the coder itself but can be generated correctly through introducing self-collaboration; and the third category includes tasks that cannot be correctly generated even with self-collaboration. By manually checking the tasks in the third category, we attribute the responsibility for generating errors to different roles and further subdivide the types of errors in Figure \ref{error_analysis}. From the findings, it is evident that the predominant error originates from the coder. This error is inherently tied to the performance of the base model. However, the introduction of self-collaboration markedly improves the quality of code generation, where 18.4\% of tasks unpassed by coder are corrected by self-collaboration. Moreover, the errors caused by the analyst mainly stem from ambiguous or incomplete requirement descriptions, which cannot be solely attributed to the analyst. Remedying this issue can be alleviated by providing more precise and complete requirement descriptions or incorporating a small amount of human guidance. On the other hand, errors associated with testers predominantly result from exception handling. These adjustments usually do not introduce new errors and tend to enhance the safety of the program. However, they may cause some instances to throw an error instead of returning a value, thus failing the test case. 

In addition to the preceding problems, the probability of deviating from the requirements caused by the additional introduction of the analyst and tester is less than 1\%, so the whole system is relatively reliable.

\begin{figure*}[h!]
\centering
\includegraphics[width=0.8\textwidth]{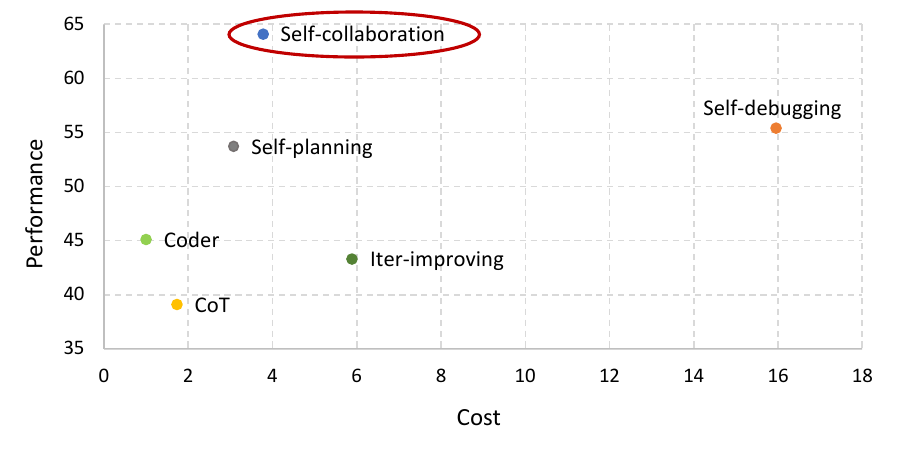}
\caption{Performance and Cost of Self-collaboration Compared to Baselines.}
\label{cost}
\end{figure*}

\subsubsection{Cost Analysis}
As illustrated in Figure \ref{cost}, we measure the cost (prompt tokens + generated tokens) and performance of self-collaboration, coder, and other prompting approaches, where we normalize the cost of coder as 1. The experimental results indicate that the improvement of the self-collaboration approach is significant, and its token usage is moderate among all prompting approaches. However, considering the high labor expenses of software development teams, the cost-effectiveness of self-collaboration is obvious.

\subsection{RQ7: Case Study}
We conduct case studies to qualitatively evaluate the performance of our approach. Our study mainly focuses on two kinds of cases. The first is a relatively simple requirement for function-level code generation, which aims to demonstrate the workflow of our virtual team. The second involves a more complex real-world requirement, specifically asking model to create a project in Python, which is a high-level requirement similar to a user story. The extended case study can be found in Appendix \ref{a_extend_case}.

\subsubsection{Workflow}
\begin{figure*}[t!]
\centering
\includegraphics[width=\textwidth]{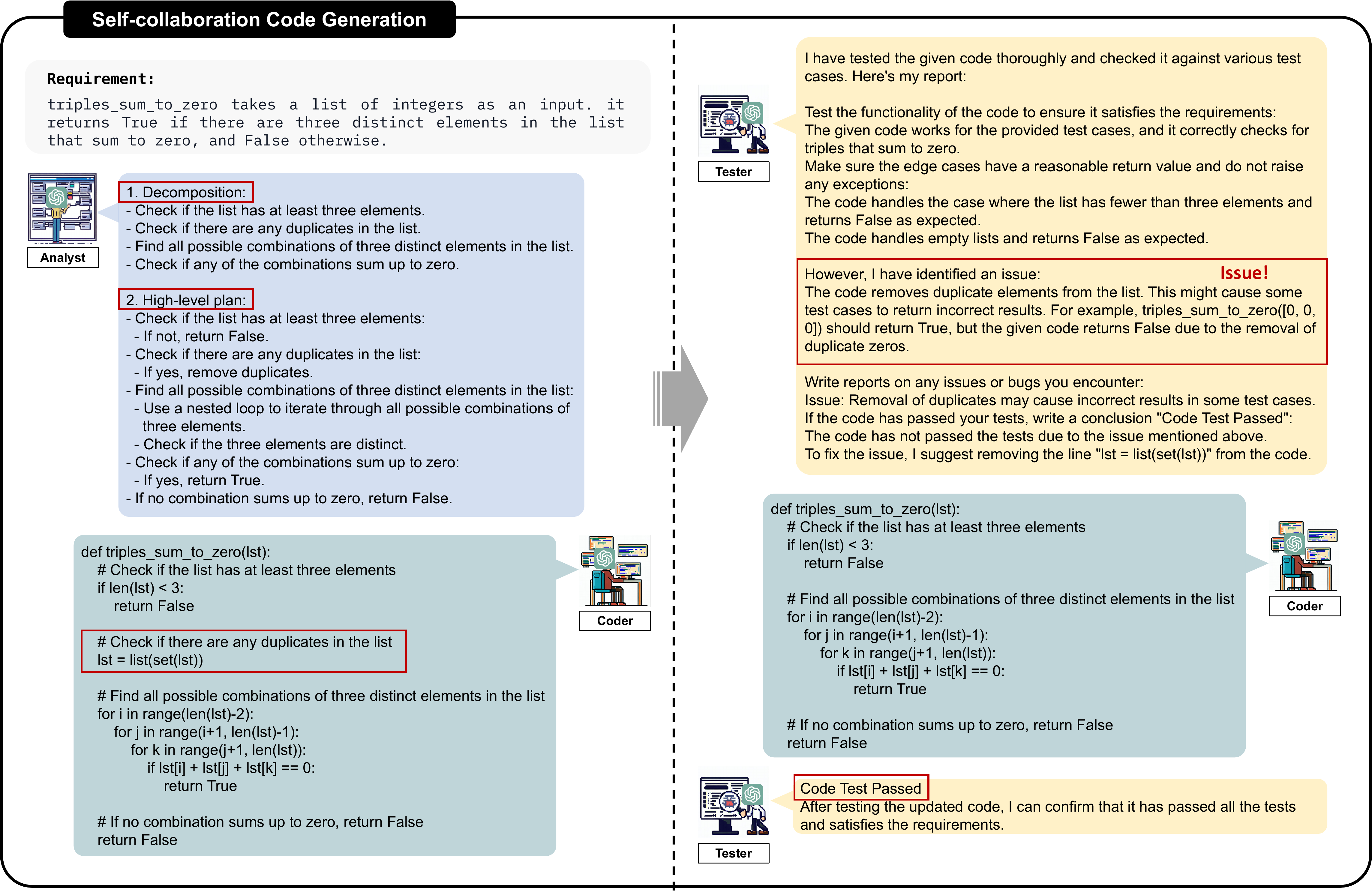}
\caption{Case study on HumanEval benchmark.}
\label{case}
\end{figure*}

We illustrate the workflow and performance of self-collaboration on HumaneEval benchmark. A case study is presented in Fig. \ref{case}, which exemplifies how the three roles work collaboratively to solve code generation in self-collaboration. 1. The analyst conducts a comprehensive analysis and devises a plan to address the entire set of requirements. For this requirement, the analyst first decomposes it into several easy-to-solve subtasks and then gives some high-level plans based on the subtasks. 2. The coder implements a code according to the given requirements, the devised decompositions, and high-level plans. It is evident that each sub-function and step within this implemented code exhibits a nearly one-to-one correspondence with decompositions and high-level plans. 3. The tester composes a detailed test report for the implemented code, identifying errors within it. In this test report, the tester thinks that the implemented code may lead to the removal of duplicate elements from the list, potentially resulting in the failure of some edge test cases. Consequently, it is recommended to eliminate the line "lst = list(set(lst))" from the implemented code. 4. The coder subsequently refines the code based on the feedback provided in the test report. In the revised code, the recommendation from the test report is taken into account, and the line "lst = list(set(lst))" is concurrently removed. 5. The tester assessed the revised code, confirming that no issues exist and all tests pass, thus concluding the code generation process. 

\subsubsection{Complex Task}
\begin{figure*}[t!]
\centering
\includegraphics[width=\textwidth]{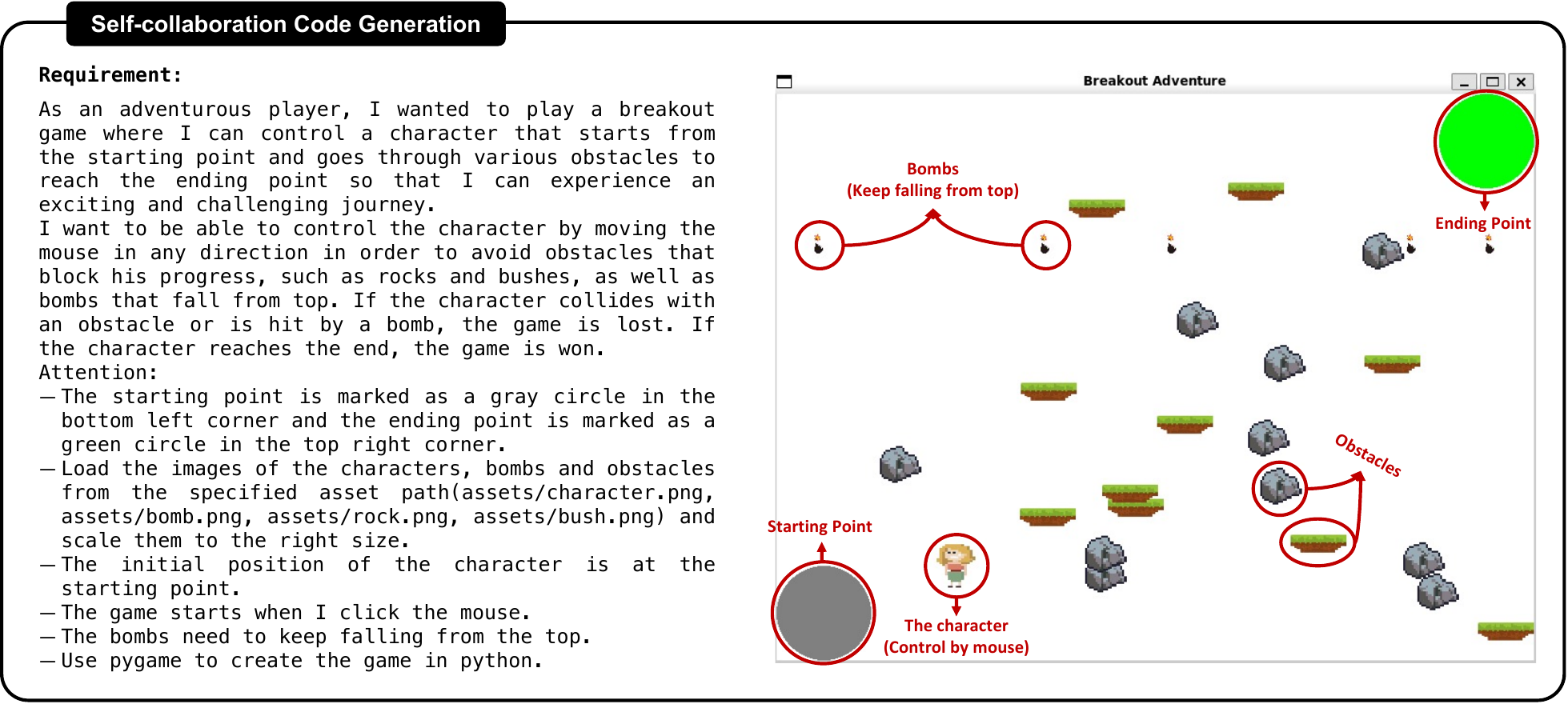}
\caption{Case study on complex tasks in real-world scenarios. Red markers are added to denote specific objects.}
\label{complex_case}
\end{figure*}

We apply self-collaboration code generation to a complex game development requirement, with a case study shown in Fig. \ref{complex_case}. Self-collaboration approach generates a working game that fulfills all requirements without human intervention. First, our approach implements the correct game logic, where the mouse clicks to start the game, and the mouse controls the character from the starting point, through all obstacles, and avoiding falling bombs to reach the ending point. The game design ensures precise character control, feedback for win/loss conditions, and correct collision detection with obstacles\&bombs. Second, the specifications outlined in the requirements have been adhered to, including the visual representation of the start and end points, the necessary game assets loading, and the proper scaling of images. Third, it also notices some game logic that is not mentioned in the requirements, but perfectly in line with common sense, such as ``Bombs fall from the top of the screen and reset their position when they hit the bottom". In contrast, the result of direct generation is a rough draft of the Python script. This script does not include all the functionality requested in the requirements. Even if we manually input the instruction ``Continue to add functionality" until ChatGPT thinks all functionality is covered, it still fails to satisfactorily fulfill this requirement actually. 

\begin{figure*}[t!]
\centering
\includegraphics[width=0.98\textwidth]{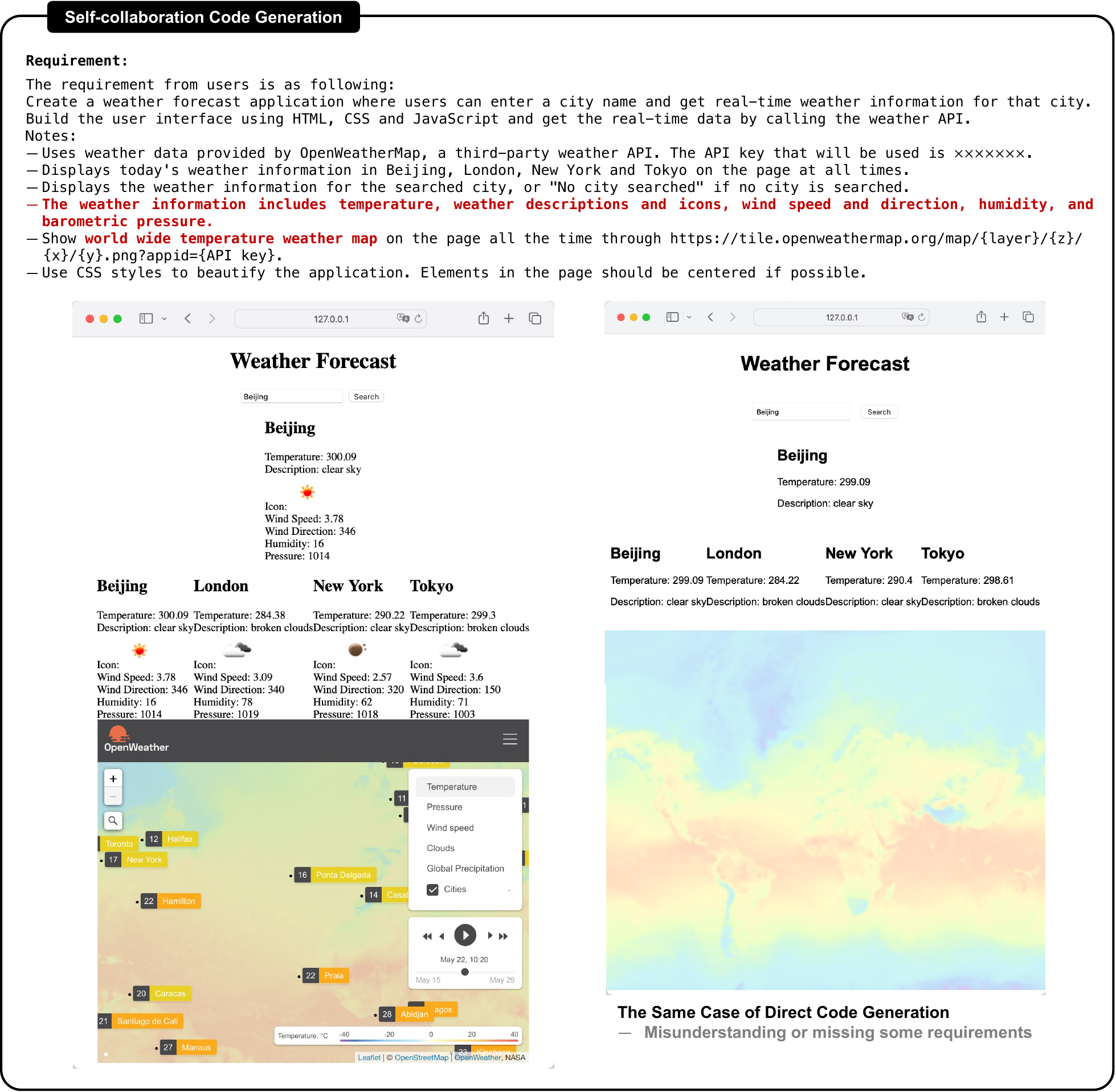}
\caption{Case study on complex tasks in real-world scenarios. }
\label{complex_case2}
\end{figure*}

We also conduct a case study related to website development. The requirement of this case requires the model to develop a weather forecast website, involving the generation of multiple types of files (HTML, CSS, Javascript), which can be considered as a micro development project. The case study of the website development is shown in Fig. \ref{complex_case2}. Our self-collaboration code generation approach produces a website that is superior to ChatGPT direct generation approach in terms of functionality and visual appeal. The analysts make the generated code comprehensive considering every requirement, including search functionality, weather information display, temperature weather map, etc. The testers assure that each requirement has not been misunderstood and is truly in line with the user's intent. In contrast, the direct generation approach occasionally falls short, either by missing certain requirements or misunderstanding them. For instance, it often neglects to include some weather information and fails to display worldwide temperature weather maps.

\section{Related Work}
In this section, we outline the most relevant directions and associated papers of this work to highlight our research's innovative and advanced nature.

\subsection{Multi-agent Collaberation}
Multi-agent collaboration refers to the coordination and collaboration among multiple artificial intelligence (AI) systems or the symbiotic collaboration between AI and human, working together to achieve a shared objective \cite{Smoliar91a}. This direction has been under exploration for a considerable length of time \cite{ClausB98,minsky2007emotion}. Recent developments indicate a promising trajectory wherein multi-agent collaboration techniques are being utilized to transcend the limitations of LLMs. The ways of multi-agent collaboration with LLMs are diverse. Recently, VisualGPT \cite{wu2023visual} and HuggingGPT \cite{shen2023hugginggpt} explore the collaboration of LLMs with other models, specifically the LLMs are used as decision centers to control and invoke other models to handle more domains, such as vision, speech, and signals. CAMEL \cite{li2023camel} explores the possibility of interaction between two LLMs. These studies primarily employ case studies in the experimental phase to demonstrate their effectiveness with specific prompts for each case. In contrast, our research utilizes both quantitative and qualitative analysis to provide a comprehensive evaluation of self-collaboration code generation. By introducing software-development methodology, we achieve steady and significant improvements for all evaluations in this paper with uniform role instructions.

Our work aligns with the concept of self-collaboration of LLMs, but it is uniquely positioned within the field of software engineering, which intrinsically fosters collaborative attributes. To the best of our knowledge, we are the first to introduce the waterfall model in software-development methodology to the collaboration among LLMs. We demonstrate the feasibility of this approach and provide the direction for future work on multi-agent collaboration.  

\subsection{Instruction and Prompt Engineering}
The success of Large Language Models (LLMs) has led to extensive attention to instruction and prompt engineering. This direction is primarily concerned with designing and optimizing model inputs, i.e., prompts or instructions, to more precisely guide the models to generate the desired outputs. This is a critical task since the behavior of large language models depends heavily on the inputs they receive \cite{GPT3, LiuYFJHN23, PACE}.

Recent studies have shown that well-designed prompting methods, such as chain-of-thought prompting (CoT) \cite{wei2022chain} and zero-shot CoT \cite{KojimaGRMI22}, can significantly improve the performance of LLMs, even beyond the scaling law. In addition, improved approaches such as Least-to-most prompting and PAL, based on the CoT approach, further enhance LLMs in solving complex tasks, such as reasoning and mathematical tasks. Meanwhile, since hand-crafted instructions or prompts may not always be optimal, some researchers have explored automated instruction or prompt generation approaches \cite{ShinRLWS20,ReynoldsM21,JiangXAN20,prompt_engineer}. These approaches try to find or optimize the best instructions or prompts for a specific task in order to improve model performance and accuracy.

In this work, we incorporate role-playing in the instructions to enable the same LLM to be differentiated into different roles, and demonstrate the effectiveness of role-playing. This encourages the model to think and problem-solve from the perspective of the given role, thus ensuring diverse and collaborative contributions toward the solution.

\subsection{Application of LLMs in Various Stages of Software Development}
With the increasing abilities of LLMs, there is an increasing interest in using them for tasks that facilitate various stages of software development, such as code generation, automated test generation, and automated program repair (APR). 

The work \cite{arora2023advancing} explores the potential of LLMs in driving requirements engineering processes. The work \cite{Parsel} applied LLMs to code generation and demonstrated significant improvement in the pass rate of generated complex programs. The work \cite{Testers} employed LLMs for generating tests to reproduce a given bug report and found that this approach holds great potential in enhancing developer efficiency. The work \cite{schafer2023empirical} also demonstrates the effectiveness of applying LLMs to test generation. In a comprehensive study conducted by the work \cite{APR2}, the direct application of LLMs for APR was explored and it was shown that LLMs outperform all existing APR techniques by a substantial margin. Additionally, the work \cite{con_APR} successfully implemented conversational APR using ChatGPT.

The applications of LLMs in software development, as highlighted above, have shown numerous successful outcomes in different stages. However, these successes are limited to individual task (stage) of software development. These tasks can be performed synergistically through LLMs to maximize their overall impact and thus achieve a higher level of automation in software development.

\subsection{Code Generation with Other Stage of Software Development}
There are a few works that use various stages of software development to enhance code generation based on LLMs without retraining. Some existing approaches pick the correct or best solution from multiple candidates by reranking. CodeT \cite{codet} employs a language model to automatically generate test cases for code samples and subsequently utilizes a dual execution agreement to rank the code. Coder-Reviewer \cite{CoderReviewer} introduces an additional reviewer model to enhance the coder model, which generates requirements based on the code and calculates the maximum likelihood to rank the generated code. They provide no feedback to the model, so that falls under the category of post-processing. Recently, another type of approach focuses on improving the quality of generated code. Self-planning \cite{Self-planning} introduces a planning stage prior to code generation. Self-debugging \cite{debug} teaches LLM to perform rubber duck debugging after code generation. They rely on few-shot prompting to instruct LLM, which requires a certain amount of effort to construct demonstration examples as prompts for each dataset.

Our work proposes self-collaboration code generation, which can cover any stage of the software development process with little effort to customize the role instructions. Each role can be regarded as an `expert' responsible for a distinct stage and provides feedback to each other to improve the quality of the final generated code. Moreover, we introduce software-development methodology, which is imperative for solving complex code generation tasks and practical software development problems.

\section{Discussion and Future Work}
In this section, we discuss some limitations of this work and present a set of potential directions for future research.

1) There are still some gaps between the evaluation benchmark we used and the actual software development scenarios. Because there is still a lack of benchmarks and corresponding evaluation metrics that can truly reflect actual software development tasks. In order to alleviate this problem, we have selected some more complex tasks originating from actual scenarios for the case study.

2) Although the virtual team we have assembled is a fully autonomous system and the roles in the virtual team can monitor and correct each other, incorporating a small amount of guidance from human experts to oversee the functioning of the virtual team would help enhance the utility of our approach in real-world application scenarios.

3) In this study, we constructed a virtual development team consisting of three roles to target the code generation task. However, for other problems, the team structure may need to be adjusted accordingly. It is worth emphasizing that our proposed approach is not limited to the traditional software engineering domain, and it has the potential to build virtual teams consisting of specific roles for a variety of problems in other domains as well.

\section{Conclusion}
In this paper, we have proposed a self-collaboration framework designed to enhance the problem-solving capability of LLMs in a collaborative and interactive way. We investigate the potential of LLMs in facilitating collaborative code generation within software development processes. Specifically, based on the proposed framework, we assemble an elementary team consisting of three distinct LLM agents, designed to address code generation tasks collaboratively. Extensive experimental results demonstrate the effectiveness and generalizability of self-collaboration framework. In conclusion, self-collaboration framework provides an effective approach to automatic code generation. This innovative approach has the potential to substantially improve the quality of generated code, reduce human intervention, and accelerate the development of complex software systems. Moreover, our work can serve as a foundation for future research on multi-agent collaboration approaches in various domains and the development of more advanced and specialized virtual teams to tackle more complex tasks.

\bibliographystyle{ACM-Reference-Format}
\bibliography{ref}


\begin{thebibliography}{65}


\ifx \showCODEN    \undefined \def \showCODEN     #1{\unskip}     \fi
\ifx \showDOI      \undefined \def \showDOI       #1{#1}\fi
\ifx \showISBNx    \undefined \def \showISBNx     #1{\unskip}     \fi
\ifx \showISBNxiii \undefined \def \showISBNxiii  #1{\unskip}     \fi
\ifx \showISSN     \undefined \def \showISSN      #1{\unskip}     \fi
\ifx \showLCCN     \undefined \def \showLCCN      #1{\unskip}     \fi
\ifx \shownote     \undefined \def \shownote      #1{#1}          \fi
\ifx \showarticletitle \undefined \def \showarticletitle #1{#1}   \fi
\ifx \showURL      \undefined \def \showURL       {\relax}        \fi
\providecommand\bibfield[2]{#2}
\providecommand\bibinfo[2]{#2}
\providecommand\natexlab[1]{#1}
\providecommand\showeprint[2][]{arXiv:#2}

\bibitem[Abrahamsson et~al\mbox{.}(2002)]%
        {abrahamsson2002agile}
\bibfield{author}{\bibinfo{person}{Pekka Abrahamsson}, \bibinfo{person}{Outi
  Salo}, \bibinfo{person}{Jussi Ronkainen}, {and} \bibinfo{person}{Juhani
  Warsta}.} \bibinfo{year}{2002}\natexlab{}.
\newblock \showarticletitle{Agile software development methods: Review and
  analysis}.
\newblock  (\bibinfo{year}{2002}).
\newblock


\bibitem[Arora et~al\mbox{.}(2023)]%
        {arora2023advancing}
\bibfield{author}{\bibinfo{person}{Chetan Arora}, \bibinfo{person}{John
  Grundy}, {and} \bibinfo{person}{Mohamed Abdelrazek}.}
  \bibinfo{year}{2023}\natexlab{}.
\newblock \showarticletitle{Advancing Requirements Engineering through
  Generative {AI:} Assessing the Role of LLMs}.
\newblock \bibinfo{journal}{\emph{CoRR}}  \bibinfo{volume}{abs/2310.13976}
  (\bibinfo{year}{2023}).
\newblock


\bibitem[Austin et~al\mbox{.}(2021)]%
        {mbpp}
\bibfield{author}{\bibinfo{person}{Jacob Austin}, \bibinfo{person}{Augustus
  Odena}, \bibinfo{person}{Maxwell~I. Nye}, \bibinfo{person}{Maarten Bosma},
  \bibinfo{person}{Henryk Michalewski}, \bibinfo{person}{David Dohan},
  \bibinfo{person}{Ellen Jiang}, \bibinfo{person}{Carrie~J. Cai},
  \bibinfo{person}{Michael Terry}, \bibinfo{person}{Quoc~V. Le}, {and}
  \bibinfo{person}{Charles Sutton}.} \bibinfo{year}{2021}\natexlab{}.
\newblock \showarticletitle{Program Synthesis with Large Language Models}.
\newblock \bibinfo{journal}{\emph{CoRR}}  \bibinfo{volume}{abs/2108.07732}
  (\bibinfo{year}{2021}).
\newblock


\bibitem[Beck et~al\mbox{.}(2001)]%
        {beck2001manifesto}
\bibfield{author}{\bibinfo{person}{Kent Beck}, \bibinfo{person}{Mike Beedle},
  \bibinfo{person}{Arie Van~Bennekum}, \bibinfo{person}{Alistair Cockburn},
  \bibinfo{person}{Ward Cunningham}, \bibinfo{person}{Martin Fowler},
  \bibinfo{person}{James Grenning}, \bibinfo{person}{Jim Highsmith},
  \bibinfo{person}{Andrew Hunt}, \bibinfo{person}{Ron Jeffries},
  {et~al\mbox{.}}} \bibinfo{year}{2001}\natexlab{}.
\newblock \showarticletitle{Manifesto for agile software development}.
\newblock  (\bibinfo{year}{2001}).
\newblock


\bibitem[Belbin(2012)]%
        {belbin2012team}
\bibfield{author}{\bibinfo{person}{R~Meredith Belbin}.}
  \bibinfo{year}{2012}\natexlab{}.
\newblock \bibinfo{booktitle}{\emph{Team roles at work}}.
\newblock \bibinfo{publisher}{Routledge}.
\newblock


\bibitem[Brown et~al\mbox{.}(2020)]%
        {GPT3}
\bibfield{author}{\bibinfo{person}{Tom~B. Brown}, \bibinfo{person}{Benjamin
  Mann}, \bibinfo{person}{Nick Ryder}, \bibinfo{person}{Melanie Subbiah},
  \bibinfo{person}{Jared Kaplan}, \bibinfo{person}{Prafulla Dhariwal},
  \bibinfo{person}{Arvind Neelakantan}, \bibinfo{person}{Pranav Shyam},
  \bibinfo{person}{Girish Sastry}, \bibinfo{person}{Amanda Askell},
  \bibinfo{person}{Sandhini Agarwal}, \bibinfo{person}{Ariel Herbert{-}Voss},
  \bibinfo{person}{Gretchen Krueger}, \bibinfo{person}{Tom Henighan},
  \bibinfo{person}{Rewon Child}, \bibinfo{person}{Aditya Ramesh},
  \bibinfo{person}{Daniel~M. Ziegler}, \bibinfo{person}{Jeffrey Wu},
  \bibinfo{person}{Clemens Winter}, \bibinfo{person}{Christopher Hesse},
  \bibinfo{person}{Mark Chen}, \bibinfo{person}{Eric Sigler},
  \bibinfo{person}{Mateusz Litwin}, \bibinfo{person}{Scott Gray},
  \bibinfo{person}{Benjamin Chess}, \bibinfo{person}{Jack Clark},
  \bibinfo{person}{Christopher Berner}, \bibinfo{person}{Sam McCandlish},
  \bibinfo{person}{Alec Radford}, \bibinfo{person}{Ilya Sutskever}, {and}
  \bibinfo{person}{Dario Amodei}.} \bibinfo{year}{2020}\natexlab{}.
\newblock \showarticletitle{Language Models are Few-Shot Learners}. In
  \bibinfo{booktitle}{\emph{NeurIPS 2020}}.
\newblock


\bibitem[Chen et~al\mbox{.}(2022)]%
        {codet}
\bibfield{author}{\bibinfo{person}{Bei Chen}, \bibinfo{person}{Fengji Zhang},
  \bibinfo{person}{Anh Nguyen}, \bibinfo{person}{Daoguang Zan},
  \bibinfo{person}{Zeqi Lin}, \bibinfo{person}{Jian{-}Guang Lou}, {and}
  \bibinfo{person}{Weizhu Chen}.} \bibinfo{year}{2022}\natexlab{}.
\newblock \showarticletitle{CodeT: Code Generation with Generated Tests}.
\newblock \bibinfo{journal}{\emph{CoRR}}  \bibinfo{volume}{abs/2207.10397}
  (\bibinfo{year}{2022}).
\newblock


\bibitem[Chen et~al\mbox{.}(2021)]%
        {codex}
\bibfield{author}{\bibinfo{person}{Mark Chen}, \bibinfo{person}{Jerry Tworek},
  \bibinfo{person}{Heewoo Jun}, \bibinfo{person}{Qiming Yuan},
  \bibinfo{person}{Henrique~Pond{\'{e}} de Oliveira~Pinto},
  \bibinfo{person}{Jared Kaplan}, \bibinfo{person}{Harrison Edwards},
  \bibinfo{person}{Yuri Burda}, \bibinfo{person}{Nicholas Joseph},
  \bibinfo{person}{Greg Brockman}, \bibinfo{person}{Alex Ray},
  \bibinfo{person}{Raul Puri}, \bibinfo{person}{Gretchen Krueger},
  \bibinfo{person}{Michael Petrov}, \bibinfo{person}{Heidy Khlaaf},
  \bibinfo{person}{Girish Sastry}, \bibinfo{person}{Pamela Mishkin},
  \bibinfo{person}{Brooke Chan}, \bibinfo{person}{Scott Gray},
  \bibinfo{person}{Nick Ryder}, \bibinfo{person}{Mikhail Pavlov},
  \bibinfo{person}{Alethea Power}, \bibinfo{person}{Lukasz Kaiser},
  \bibinfo{person}{Mohammad Bavarian}, \bibinfo{person}{Clemens Winter},
  \bibinfo{person}{Philippe Tillet}, \bibinfo{person}{Felipe~Petroski Such},
  \bibinfo{person}{Dave Cummings}, \bibinfo{person}{Matthias Plappert},
  \bibinfo{person}{Fotios Chantzis}, \bibinfo{person}{Elizabeth Barnes},
  \bibinfo{person}{Ariel Herbert{-}Voss}, \bibinfo{person}{William~Hebgen
  Guss}, \bibinfo{person}{Alex Nichol}, \bibinfo{person}{Alex Paino},
  \bibinfo{person}{Nikolas Tezak}, \bibinfo{person}{Jie Tang},
  \bibinfo{person}{Igor Babuschkin}, \bibinfo{person}{Suchir Balaji},
  \bibinfo{person}{Shantanu Jain}, \bibinfo{person}{William Saunders},
  \bibinfo{person}{Christopher Hesse}, \bibinfo{person}{Andrew~N. Carr},
  \bibinfo{person}{Jan Leike}, \bibinfo{person}{Joshua Achiam},
  \bibinfo{person}{Vedant Misra}, \bibinfo{person}{Evan Morikawa},
  \bibinfo{person}{Alec Radford}, \bibinfo{person}{Matthew Knight},
  \bibinfo{person}{Miles Brundage}, \bibinfo{person}{Mira Murati},
  \bibinfo{person}{Katie Mayer}, \bibinfo{person}{Peter Welinder},
  \bibinfo{person}{Bob McGrew}, \bibinfo{person}{Dario Amodei},
  \bibinfo{person}{Sam McCandlish}, \bibinfo{person}{Ilya Sutskever}, {and}
  \bibinfo{person}{Wojciech Zaremba}.} \bibinfo{year}{2021}\natexlab{}.
\newblock \showarticletitle{Evaluating Large Language Models Trained on Code}.
\newblock \bibinfo{journal}{\emph{CoRR}} (\bibinfo{year}{2021}).
\newblock
\urldef\tempurl%
\url{https://arxiv.org/abs/2107.03374}
\showURL{%
\tempurl}


\bibitem[Chen et~al\mbox{.}(2023)]%
        {debug}
\bibfield{author}{\bibinfo{person}{Xinyun Chen}, \bibinfo{person}{Maxwell Lin},
  \bibinfo{person}{Nathanael Sch{\"{a}}rli}, {and} \bibinfo{person}{Denny
  Zhou}.} \bibinfo{year}{2023}\natexlab{}.
\newblock \showarticletitle{Teaching Large Language Models to Self-Debug}.
\newblock \bibinfo{journal}{\emph{CoRR}}  \bibinfo{volume}{abs/2304.05128}
  (\bibinfo{year}{2023}).
\newblock


\bibitem[Chowdhery et~al\mbox{.}(2022)]%
        {PaLM}
\bibfield{author}{\bibinfo{person}{Aakanksha Chowdhery},
  \bibinfo{person}{Sharan Narang}, \bibinfo{person}{Jacob Devlin},
  \bibinfo{person}{Maarten Bosma}, \bibinfo{person}{Gaurav Mishra},
  \bibinfo{person}{Adam Roberts}, \bibinfo{person}{Paul Barham},
  \bibinfo{person}{Hyung~Won Chung}, \bibinfo{person}{Charles Sutton},
  \bibinfo{person}{Sebastian Gehrmann}, \bibinfo{person}{Parker Schuh},
  \bibinfo{person}{Kensen Shi}, \bibinfo{person}{Sasha Tsvyashchenko},
  \bibinfo{person}{Joshua Maynez}, \bibinfo{person}{Abhishek Rao},
  \bibinfo{person}{Parker Barnes}, \bibinfo{person}{Yi Tay},
  \bibinfo{person}{Noam Shazeer}, \bibinfo{person}{Vinodkumar Prabhakaran},
  \bibinfo{person}{Emily Reif}, \bibinfo{person}{Nan Du}, \bibinfo{person}{Ben
  Hutchinson}, \bibinfo{person}{Reiner Pope}, \bibinfo{person}{James Bradbury},
  \bibinfo{person}{Jacob Austin}, \bibinfo{person}{Michael Isard},
  \bibinfo{person}{Guy Gur{-}Ari}, \bibinfo{person}{Pengcheng Yin},
  \bibinfo{person}{Toju Duke}, \bibinfo{person}{Anselm Levskaya},
  \bibinfo{person}{Sanjay Ghemawat}, \bibinfo{person}{Sunipa Dev},
  \bibinfo{person}{Henryk Michalewski}, \bibinfo{person}{Xavier Garcia},
  \bibinfo{person}{Vedant Misra}, \bibinfo{person}{Kevin Robinson},
  \bibinfo{person}{Liam Fedus}, \bibinfo{person}{Denny Zhou},
  \bibinfo{person}{Daphne Ippolito}, \bibinfo{person}{David Luan},
  \bibinfo{person}{Hyeontaek Lim}, \bibinfo{person}{Barret Zoph},
  \bibinfo{person}{Alexander Spiridonov}, \bibinfo{person}{Ryan Sepassi},
  \bibinfo{person}{David Dohan}, \bibinfo{person}{Shivani Agrawal},
  \bibinfo{person}{Mark Omernick}, \bibinfo{person}{Andrew~M. Dai},
  \bibinfo{person}{Thanumalayan~Sankaranarayana Pillai}, \bibinfo{person}{Marie
  Pellat}, \bibinfo{person}{Aitor Lewkowycz}, \bibinfo{person}{Erica Moreira},
  \bibinfo{person}{Rewon Child}, \bibinfo{person}{Oleksandr Polozov},
  \bibinfo{person}{Katherine Lee}, \bibinfo{person}{Zongwei Zhou},
  \bibinfo{person}{Xuezhi Wang}, \bibinfo{person}{Brennan Saeta},
  \bibinfo{person}{Mark Diaz}, \bibinfo{person}{Orhan Firat},
  \bibinfo{person}{Michele Catasta}, \bibinfo{person}{Jason Wei},
  \bibinfo{person}{Kathy Meier{-}Hellstern}, \bibinfo{person}{Douglas Eck},
  \bibinfo{person}{Jeff Dean}, \bibinfo{person}{Slav Petrov}, {and}
  \bibinfo{person}{Noah Fiedel}.} \bibinfo{year}{2022}\natexlab{}.
\newblock \showarticletitle{PaLM: Scaling Language Modeling with Pathways}.
\newblock \bibinfo{journal}{\emph{CoRR}}  \bibinfo{volume}{abs/2204.02311}
  (\bibinfo{year}{2022}).
\newblock


\bibitem[Chung et~al\mbox{.}(2022)]%
        {FlanPaLM}
\bibfield{author}{\bibinfo{person}{Hyung~Won Chung}, \bibinfo{person}{Le Hou},
  \bibinfo{person}{Shayne Longpre}, \bibinfo{person}{Barret Zoph},
  \bibinfo{person}{Yi Tay}, \bibinfo{person}{William Fedus},
  \bibinfo{person}{Eric Li}, \bibinfo{person}{Xuezhi Wang},
  \bibinfo{person}{Mostafa Dehghani}, \bibinfo{person}{Siddhartha Brahma},
  \bibinfo{person}{Albert Webson}, \bibinfo{person}{Shixiang~Shane Gu},
  \bibinfo{person}{Zhuyun Dai}, \bibinfo{person}{Mirac Suzgun},
  \bibinfo{person}{Xinyun Chen}, \bibinfo{person}{Aakanksha Chowdhery},
  \bibinfo{person}{Sharan Narang}, \bibinfo{person}{Gaurav Mishra},
  \bibinfo{person}{Adams Yu}, \bibinfo{person}{Vincent~Y. Zhao},
  \bibinfo{person}{Yanping Huang}, \bibinfo{person}{Andrew~M. Dai},
  \bibinfo{person}{Hongkun Yu}, \bibinfo{person}{Slav Petrov},
  \bibinfo{person}{Ed~H. Chi}, \bibinfo{person}{Jeff Dean},
  \bibinfo{person}{Jacob Devlin}, \bibinfo{person}{Adam Roberts},
  \bibinfo{person}{Denny Zhou}, \bibinfo{person}{Quoc~V. Le}, {and}
  \bibinfo{person}{Jason Wei}.} \bibinfo{year}{2022}\natexlab{}.
\newblock \showarticletitle{Scaling Instruction-Finetuned Language Models}.
\newblock \bibinfo{journal}{\emph{CoRR}}  \bibinfo{volume}{abs/2210.11416}
  (\bibinfo{year}{2022}).
\newblock


\bibitem[Claus and Boutilier(1998)]%
        {ClausB98}
\bibfield{author}{\bibinfo{person}{Caroline Claus} {and} \bibinfo{person}{Craig
  Boutilier}.} \bibinfo{year}{1998}\natexlab{}.
\newblock \showarticletitle{The Dynamics of Reinforcement Learning in
  Cooperative Multiagent Systems}. In \bibinfo{booktitle}{\emph{{AAAI/IAAI}}}.
  \bibinfo{publisher}{{AAAI} Press / The {MIT} Press},
  \bibinfo{pages}{746--752}.
\newblock


\bibitem[DeMarco and Lister(2013)]%
        {demarco2013peopleware}
\bibfield{author}{\bibinfo{person}{Tom DeMarco} {and} \bibinfo{person}{Tim
  Lister}.} \bibinfo{year}{2013}\natexlab{}.
\newblock \bibinfo{booktitle}{\emph{Peopleware: productive projects and
  teams}}.
\newblock \bibinfo{publisher}{Addison-Wesley}.
\newblock


\bibitem[Dong et~al\mbox{.}(2023a)]%
        {CodeScore}
\bibfield{author}{\bibinfo{person}{Yihong Dong}, \bibinfo{person}{Jiazheng
  Ding}, \bibinfo{person}{Xue Jiang}, \bibinfo{person}{Zhuo Li},
  \bibinfo{person}{Ge Li}, {and} \bibinfo{person}{Zhi Jin}.}
  \bibinfo{year}{2023}\natexlab{a}.
\newblock \showarticletitle{CodeScore: Evaluating Code Generation by Learning
  Code Execution}.
\newblock \bibinfo{journal}{\emph{CoRR}}  \bibinfo{volume}{abs/2301.09043}
  (\bibinfo{year}{2023}).
\newblock


\bibitem[Dong et~al\mbox{.}(2024)]%
        {CDD}
\bibfield{author}{\bibinfo{person}{Yihong Dong}, \bibinfo{person}{Xue Jiang},
  \bibinfo{person}{Huanyu Liu}, \bibinfo{person}{Zhi Jin}, {and}
  \bibinfo{person}{Ge Li}.} \bibinfo{year}{2024}\natexlab{}.
\newblock \showarticletitle{Generalization or Memorization: Data Contamination
  and Trustworthy Evaluation for Large Language Models}.
\newblock \bibinfo{journal}{\emph{CoRR}}  \bibinfo{volume}{abs/2402.15938}
  (\bibinfo{year}{2024}).
\newblock


\bibitem[Dong et~al\mbox{.}(2023b)]%
        {CODEP}
\bibfield{author}{\bibinfo{person}{Yihong Dong}, \bibinfo{person}{Ge Li}, {and}
  \bibinfo{person}{Zhi Jin}.} \bibinfo{year}{2023}\natexlab{b}.
\newblock \showarticletitle{{CODEP:} Grammatical Seq2Seq Model for
  General-Purpose Code Generation}. In \bibinfo{booktitle}{\emph{{ISSTA}}}.
  \bibinfo{publisher}{{ACM}}, \bibinfo{pages}{188--198}.
\newblock


\bibitem[Dong et~al\mbox{.}(2023c)]%
        {PACE}
\bibfield{author}{\bibinfo{person}{Yihong Dong}, \bibinfo{person}{Kangcheng
  Luo}, \bibinfo{person}{Xue Jiang}, \bibinfo{person}{Zhi Jin}, {and}
  \bibinfo{person}{Ge Li}.} \bibinfo{year}{2023}\natexlab{c}.
\newblock \showarticletitle{{PACE:} Improving Prompt with Actor-Critic Editing
  for Large Language Model}.
\newblock \bibinfo{journal}{\emph{CoRR}}  \bibinfo{volume}{abs/2308.10088}
  (\bibinfo{year}{2023}).
\newblock


\bibitem[Fried et~al\mbox{.}(2022)]%
        {incoder}
\bibfield{author}{\bibinfo{person}{Daniel Fried}, \bibinfo{person}{Armen
  Aghajanyan}, \bibinfo{person}{Jessy Lin}, \bibinfo{person}{Sida Wang},
  \bibinfo{person}{Eric Wallace}, \bibinfo{person}{Freda Shi},
  \bibinfo{person}{Ruiqi Zhong}, \bibinfo{person}{Wen{-}tau Yih},
  \bibinfo{person}{Luke Zettlemoyer}, {and} \bibinfo{person}{Mike Lewis}.}
  \bibinfo{year}{2022}\natexlab{}.
\newblock \showarticletitle{InCoder: {A} Generative Model for Code Infilling
  and Synthesis}.
\newblock \bibinfo{journal}{\emph{CoRR}}  \bibinfo{volume}{abs/2204.05999}
  (\bibinfo{year}{2022}).
\newblock


\bibitem[GitHub(2022)]%
        {Copilot}
\bibfield{author}{\bibinfo{person}{GitHub}.} \bibinfo{year}{2022}\natexlab{}.
\newblock \bibinfo{booktitle}{\emph{{Copilot}}}.
\newblock
\urldef\tempurl%
\url{https://github.com/features/copilot}
\showURL{%
\tempurl}


\bibitem[Hendrycks et~al\mbox{.}(2021)]%
        {APPS}
\bibfield{author}{\bibinfo{person}{Dan Hendrycks}, \bibinfo{person}{Steven
  Basart}, \bibinfo{person}{Saurav Kadavath}, \bibinfo{person}{Mantas Mazeika},
  \bibinfo{person}{Akul Arora}, \bibinfo{person}{Ethan Guo},
  \bibinfo{person}{Collin Burns}, \bibinfo{person}{Samir Puranik},
  \bibinfo{person}{Horace He}, \bibinfo{person}{Dawn Song}, {and}
  \bibinfo{person}{Jacob Steinhardt}.} \bibinfo{year}{2021}\natexlab{}.
\newblock \showarticletitle{Measuring Coding Challenge Competence With {APPS}}.
  In \bibinfo{booktitle}{\emph{NeurIPS Datasets and Benchmarks}}.
\newblock


\bibitem[Huggingface(2023)]%
        {HuggingChat}
\bibfield{author}{\bibinfo{person}{Huggingface}.}
  \bibinfo{year}{2023}\natexlab{}.
\newblock \bibinfo{booktitle}{\emph{HuggingChat}}.
\newblock
\urldef\tempurl%
\url{https://huggingface.co/chat/}
\showURL{%
\tempurl}


\bibitem[IBM(2023)]%
        {Dromedary}
\bibfield{author}{\bibinfo{person}{IBM}.} \bibinfo{year}{2023}\natexlab{}.
\newblock \bibinfo{booktitle}{\emph{Dromedary}}.
\newblock
\urldef\tempurl%
\url{https://huggingface.co/ibm/dromedary-65b-lora-delta-v0}
\showURL{%
\tempurl}


\bibitem[Jiang et~al\mbox{.}(2023)]%
        {Self-planning}
\bibfield{author}{\bibinfo{person}{Xue Jiang}, \bibinfo{person}{Yihong Dong},
  \bibinfo{person}{Lecheng Wang}, \bibinfo{person}{Qiwei Shang}, {and}
  \bibinfo{person}{Ge Li}.} \bibinfo{year}{2023}\natexlab{}.
\newblock \showarticletitle{Self-planning Code Generation with Large Language
  Model}.
\newblock \bibinfo{journal}{\emph{CoRR}}  \bibinfo{volume}{abs/2303.06689}
  (\bibinfo{year}{2023}).
\newblock


\bibitem[Jiang et~al\mbox{.}(2020)]%
        {JiangXAN20}
\bibfield{author}{\bibinfo{person}{Zhengbao Jiang}, \bibinfo{person}{Frank~F.
  Xu}, \bibinfo{person}{Jun Araki}, {and} \bibinfo{person}{Graham Neubig}.}
  \bibinfo{year}{2020}\natexlab{}.
\newblock \showarticletitle{How Can We Know What Language Models Know}.
\newblock \bibinfo{journal}{\emph{Trans. Assoc. Comput. Linguistics}}
  \bibinfo{volume}{8} (\bibinfo{year}{2020}), \bibinfo{pages}{423--438}.
\newblock


\bibitem[Kang et~al\mbox{.}(2022)]%
        {Testers}
\bibfield{author}{\bibinfo{person}{Sungmin Kang}, \bibinfo{person}{Juyeon
  Yoon}, {and} \bibinfo{person}{Shin Yoo}.} \bibinfo{year}{2022}\natexlab{}.
\newblock \showarticletitle{Large Language Models are Few-shot Testers:
  Exploring LLM-based General Bug Reproduction}.
\newblock \bibinfo{journal}{\emph{CoRR}}  \bibinfo{volume}{abs/2209.11515}
  (\bibinfo{year}{2022}).
\newblock


\bibitem[Katzenbach and Smith(2015)]%
        {katzenbach2015wisdom}
\bibfield{author}{\bibinfo{person}{Jon~R Katzenbach} {and}
  \bibinfo{person}{Douglas~K Smith}.} \bibinfo{year}{2015}\natexlab{}.
\newblock \bibinfo{booktitle}{\emph{The wisdom of teams: Creating the
  high-performance organization}}.
\newblock \bibinfo{publisher}{Harvard Business Review Press}.
\newblock


\bibitem[Kojima et~al\mbox{.}(2022)]%
        {KojimaGRMI22}
\bibfield{author}{\bibinfo{person}{Takeshi Kojima},
  \bibinfo{person}{Shixiang~Shane Gu}, \bibinfo{person}{Machel Reid},
  \bibinfo{person}{Yutaka Matsuo}, {and} \bibinfo{person}{Yusuke Iwasawa}.}
  \bibinfo{year}{2022}\natexlab{}.
\newblock \showarticletitle{Large Language Models are Zero-Shot Reasoners}. In
  \bibinfo{booktitle}{\emph{NeurIPS}}.
\newblock


\bibitem[Li et~al\mbox{.}(2023b)]%
        {li2023camel}
\bibfield{author}{\bibinfo{person}{Guohao Li}, \bibinfo{person}{Hasan Abed
  Al~Kader Hammoud}, \bibinfo{person}{Hani Itani}, \bibinfo{person}{Dmitrii
  Khizbullin}, {and} \bibinfo{person}{Bernard Ghanem}.}
  \bibinfo{year}{2023}\natexlab{b}.
\newblock \showarticletitle{CAMEL: Communicative Agents for" Mind" Exploration
  of Large Scale Language Model Society}.
\newblock \bibinfo{journal}{\emph{CoRR}}  \bibinfo{volume}{abs/2303.17760}
  (\bibinfo{year}{2023}).
\newblock


\bibitem[Li et~al\mbox{.}(2023a)]%
        {starcoder}
\bibfield{author}{\bibinfo{person}{Raymond Li}, \bibinfo{person}{Loubna~Ben
  Allal}, \bibinfo{person}{Yangtian Zi}, \bibinfo{person}{Niklas Muennighoff},
  \bibinfo{person}{Denis Kocetkov}, \bibinfo{person}{Chenghao Mou},
  \bibinfo{person}{Marc Marone}, \bibinfo{person}{Christopher Akiki},
  \bibinfo{person}{Jia Li}, \bibinfo{person}{Jenny Chim}, \bibinfo{person}{Qian
  Liu}, \bibinfo{person}{Evgenii Zheltonozhskii}, \bibinfo{person}{Terry~Yue
  Zhuo}, \bibinfo{person}{Thomas Wang}, \bibinfo{person}{Olivier Dehaene},
  \bibinfo{person}{Mishig Davaadorj}, \bibinfo{person}{Joel Lamy{-}Poirier},
  \bibinfo{person}{Jo{\~{a}}o Monteiro}, \bibinfo{person}{Oleh Shliazhko},
  \bibinfo{person}{Nicolas Gontier}, \bibinfo{person}{Nicholas Meade},
  \bibinfo{person}{Armel Zebaze}, \bibinfo{person}{Ming{-}Ho Yee},
  \bibinfo{person}{Logesh~Kumar Umapathi}, \bibinfo{person}{Jian Zhu},
  \bibinfo{person}{Benjamin Lipkin}, \bibinfo{person}{Muhtasham Oblokulov},
  \bibinfo{person}{Zhiruo Wang}, \bibinfo{person}{Rudra~Murthy V},
  \bibinfo{person}{Jason Stillerman}, \bibinfo{person}{Siva~Sankalp Patel},
  \bibinfo{person}{Dmitry Abulkhanov}, \bibinfo{person}{Marco Zocca},
  \bibinfo{person}{Manan Dey}, \bibinfo{person}{Zhihan Zhang},
  \bibinfo{person}{Nour Moustafa{-}Fahmy}, \bibinfo{person}{Urvashi
  Bhattacharyya}, \bibinfo{person}{Wenhao Yu}, \bibinfo{person}{Swayam Singh},
  \bibinfo{person}{Sasha Luccioni}, \bibinfo{person}{Paulo Villegas},
  \bibinfo{person}{Maxim Kunakov}, \bibinfo{person}{Fedor Zhdanov},
  \bibinfo{person}{Manuel Romero}, \bibinfo{person}{Tony Lee},
  \bibinfo{person}{Nadav Timor}, \bibinfo{person}{Jennifer Ding},
  \bibinfo{person}{Claire Schlesinger}, \bibinfo{person}{Hailey Schoelkopf},
  \bibinfo{person}{Jan Ebert}, \bibinfo{person}{Tri Dao},
  \bibinfo{person}{Mayank Mishra}, \bibinfo{person}{Alex Gu},
  \bibinfo{person}{Jennifer Robinson}, \bibinfo{person}{Carolyn~Jane Anderson},
  \bibinfo{person}{Brendan Dolan{-}Gavitt}, \bibinfo{person}{Danish
  Contractor}, \bibinfo{person}{Siva Reddy}, \bibinfo{person}{Daniel Fried},
  \bibinfo{person}{Dzmitry Bahdanau}, \bibinfo{person}{Yacine Jernite},
  \bibinfo{person}{Carlos~Mu{\~{n}}oz Ferrandis}, \bibinfo{person}{Sean
  Hughes}, \bibinfo{person}{Thomas Wolf}, \bibinfo{person}{Arjun Guha},
  \bibinfo{person}{Leandro von Werra}, {and} \bibinfo{person}{Harm de Vries}.}
  \bibinfo{year}{2023}\natexlab{a}.
\newblock \showarticletitle{StarCoder: may the source be with you!}
\newblock \bibinfo{journal}{\emph{CoRR}}  \bibinfo{volume}{abs/2305.06161}
  (\bibinfo{year}{2023}).
\newblock


\bibitem[Li et~al\mbox{.}(2022)]%
        {alphacode}
\bibfield{author}{\bibinfo{person}{Yujia Li}, \bibinfo{person}{David Choi},
  \bibinfo{person}{Junyoung Chung}, \bibinfo{person}{Nate Kushman},
  \bibinfo{person}{Julian Schrittwieser}, \bibinfo{person}{R{\'e}mi Leblond},
  \bibinfo{person}{Tom Eccles}, \bibinfo{person}{James Keeling},
  \bibinfo{person}{Felix Gimeno}, \bibinfo{person}{Agustin Dal~Lago},
  {et~al\mbox{.}}} \bibinfo{year}{2022}\natexlab{}.
\newblock \showarticletitle{Competition-level code generation with alphacode}.
\newblock \bibinfo{journal}{\emph{Science}} \bibinfo{volume}{378},
  \bibinfo{number}{6624} (\bibinfo{year}{2022}), \bibinfo{pages}{1092--1097}.
\newblock


\bibitem[Liu et~al\mbox{.}(2023)]%
        {LiuYFJHN23}
\bibfield{author}{\bibinfo{person}{Pengfei Liu}, \bibinfo{person}{Weizhe Yuan},
  \bibinfo{person}{Jinlan Fu}, \bibinfo{person}{Zhengbao Jiang},
  \bibinfo{person}{Hiroaki Hayashi}, {and} \bibinfo{person}{Graham Neubig}.}
  \bibinfo{year}{2023}\natexlab{}.
\newblock \showarticletitle{Pre-train, Prompt, and Predict: {A} Systematic
  Survey of Prompting Methods in Natural Language Processing}.
\newblock \bibinfo{journal}{\emph{{ACM} Comput. Surv.}} \bibinfo{volume}{55},
  \bibinfo{number}{9} (\bibinfo{year}{2023}), \bibinfo{pages}{195:1--195:35}.
\newblock


\bibitem[LMSYS(2023a)]%
        {Fastchat}
\bibfield{author}{\bibinfo{person}{LMSYS}.} \bibinfo{year}{2023}\natexlab{a}.
\newblock \bibinfo{booktitle}{\emph{{Fastchat}}}.
\newblock
\urldef\tempurl%
\url{https://huggingface.co/lmsys/fastchat-t5-3b-v1.0}
\showURL{%
\tempurl}


\bibitem[LMSYS(2023b)]%
        {Vicuna}
\bibfield{author}{\bibinfo{person}{LMSYS}.} \bibinfo{year}{2023}\natexlab{b}.
\newblock \bibinfo{booktitle}{\emph{Vicuna: An Open-Source Chatbot Impressing
  GPT-4 with 90\%* ChatGPT Quality}}.
\newblock
\urldef\tempurl%
\url{https://lmsys.org/blog/2023-03-30-vicuna/}
\showURL{%
\tempurl}


\bibitem[McChesney and Gallagher(2004)]%
        {mcchesney2004communication}
\bibfield{author}{\bibinfo{person}{Ian~R McChesney} {and}
  \bibinfo{person}{Seamus Gallagher}.} \bibinfo{year}{2004}\natexlab{}.
\newblock \showarticletitle{Communication and co-ordination practices in
  software engineering projects}.
\newblock \bibinfo{journal}{\emph{Information and Software Technology}}
  \bibinfo{volume}{46}, \bibinfo{number}{7} (\bibinfo{year}{2004}),
  \bibinfo{pages}{473--489}.
\newblock


\bibitem[Minsky(2007)]%
        {minsky2007emotion}
\bibfield{author}{\bibinfo{person}{Marvin Minsky}.}
  \bibinfo{year}{2007}\natexlab{}.
\newblock \bibinfo{booktitle}{\emph{The emotion machine: Commonsense thinking,
  artificial intelligence, and the future of the human mind}}.
\newblock \bibinfo{publisher}{Simon and Schuster}.
\newblock


\bibitem[MosaicML(2023)]%
        {MosaicML2023Introducing}
\bibfield{author}{\bibinfo{person}{MosaicML}.} \bibinfo{year}{2023}\natexlab{}.
\newblock \bibinfo{booktitle}{\emph{Introducing MPT-7B: A New Standard for
  Open-Source, ly Usable LLMs}}.
\newblock
\urldef\tempurl%
\url{www.mosaicml.com/blog/mpt-7b}
\showURL{%
\tempurl}


\bibitem[Nii(1986)]%
        {Blackboard}
\bibfield{author}{\bibinfo{person}{H~Penny Nii}.}
  \bibinfo{year}{1986}\natexlab{}.
\newblock \showarticletitle{Blackboard Systems.}
\newblock  (\bibinfo{year}{1986}).
\newblock


\bibitem[Nijkamp et~al\mbox{.}(2022)]%
        {nijkamp2022codegen}
\bibfield{author}{\bibinfo{person}{Erik Nijkamp}, \bibinfo{person}{Bo Pang},
  \bibinfo{person}{Hiroaki Hayashi}, \bibinfo{person}{Lifu Tu},
  \bibinfo{person}{Huan Wang}, \bibinfo{person}{Yingbo Zhou},
  \bibinfo{person}{Silvio Savarese}, {and} \bibinfo{person}{Caiming Xiong}.}
  \bibinfo{year}{2022}\natexlab{}.
\newblock \showarticletitle{Codegen: An open large language model for code with
  multi-turn program synthesis}.
\newblock \bibinfo{journal}{\emph{CoRR}}  \bibinfo{volume}{abs/2203.13474}
  (\bibinfo{year}{2022}).
\newblock


\bibitem[OpenAI(2022)]%
        {ChatGPT}
\bibfield{author}{\bibinfo{person}{OpenAI}.} \bibinfo{year}{2022}\natexlab{}.
\newblock \bibinfo{booktitle}{\emph{{ChatGPT}}}.
\newblock
\urldef\tempurl%
\url{https://openai.com/blog/chatgpt/}
\showURL{%
\tempurl}


\bibitem[OpenAI(2023)]%
        {GPT-4}
\bibfield{author}{\bibinfo{person}{OpenAI}.} \bibinfo{year}{2023}\natexlab{}.
\newblock \showarticletitle{{GPT-4} Technical Report}.
\newblock \bibinfo{journal}{\emph{CoRR}}  \bibinfo{volume}{abs/2303.08774}
  (\bibinfo{year}{2023}).
\newblock


\bibitem[Ouyang et~al\mbox{.}(2022a)]%
        {instruction}
\bibfield{author}{\bibinfo{person}{Long Ouyang}, \bibinfo{person}{Jeff Wu},
  \bibinfo{person}{Xu Jiang}, \bibinfo{person}{Diogo Almeida},
  \bibinfo{person}{Carroll~L. Wainwright}, \bibinfo{person}{Pamela Mishkin},
  \bibinfo{person}{Chong Zhang}, \bibinfo{person}{Sandhini Agarwal},
  \bibinfo{person}{Katarina Slama}, \bibinfo{person}{Alex Ray},
  \bibinfo{person}{John Schulman}, \bibinfo{person}{Jacob Hilton},
  \bibinfo{person}{Fraser Kelton}, \bibinfo{person}{Luke Miller},
  \bibinfo{person}{Maddie Simens}, \bibinfo{person}{Amanda Askell},
  \bibinfo{person}{Peter Welinder}, \bibinfo{person}{Paul~F. Christiano},
  \bibinfo{person}{Jan Leike}, {and} \bibinfo{person}{Ryan Lowe}.}
  \bibinfo{year}{2022}\natexlab{a}.
\newblock \showarticletitle{Training language models to follow instructions
  with human feedback}.
\newblock \bibinfo{journal}{\emph{CoRR}}  \bibinfo{volume}{abs/2203.02155}
  (\bibinfo{year}{2022}).
\newblock


\bibitem[Ouyang et~al\mbox{.}(2022b)]%
        {InstructGPT}
\bibfield{author}{\bibinfo{person}{Long Ouyang}, \bibinfo{person}{Jeff Wu},
  \bibinfo{person}{Xu Jiang}, \bibinfo{person}{Diogo Almeida},
  \bibinfo{person}{Carroll~L. Wainwright}, \bibinfo{person}{Pamela Mishkin},
  \bibinfo{person}{Chong Zhang}, \bibinfo{person}{Sandhini Agarwal},
  \bibinfo{person}{Katarina Slama}, \bibinfo{person}{Alex Ray},
  \bibinfo{person}{John Schulman}, \bibinfo{person}{Jacob Hilton},
  \bibinfo{person}{Fraser Kelton}, \bibinfo{person}{Luke Miller},
  \bibinfo{person}{Maddie Simens}, \bibinfo{person}{Amanda Askell},
  \bibinfo{person}{Peter Welinder}, \bibinfo{person}{Paul~F. Christiano},
  \bibinfo{person}{Jan Leike}, {and} \bibinfo{person}{Ryan Lowe}.}
  \bibinfo{year}{2022}\natexlab{b}.
\newblock \showarticletitle{Training language models to follow instructions
  with human feedback}.
\newblock \bibinfo{journal}{\emph{CoRR}}  \bibinfo{volume}{abs/2203.02155}
  (\bibinfo{year}{2022}).
\newblock


\bibitem[Petersen et~al\mbox{.}(2009)]%
        {Waterfall}
\bibfield{author}{\bibinfo{person}{Kai Petersen}, \bibinfo{person}{Claes
  Wohlin}, {and} \bibinfo{person}{Dejan Baca}.}
  \bibinfo{year}{2009}\natexlab{}.
\newblock \showarticletitle{The Waterfall Model in Large-Scale Development}. In
  \bibinfo{booktitle}{\emph{{PROFES}}} \emph{(\bibinfo{series}{Lecture years in
  Business Information Processing}, Vol.~\bibinfo{volume}{32})}.
  \bibinfo{publisher}{Springer}, \bibinfo{pages}{386--400}.
\newblock


\bibitem[Reynolds and McDonell(2021)]%
        {ReynoldsM21}
\bibfield{author}{\bibinfo{person}{Laria Reynolds} {and} \bibinfo{person}{Kyle
  McDonell}.} \bibinfo{year}{2021}\natexlab{}.
\newblock \showarticletitle{Prompt Programming for Large Language Models:
  Beyond the Few-Shot Paradigm}. In \bibinfo{booktitle}{\emph{{CHI} Extended
  Abstracts}}. \bibinfo{publisher}{{ACM}}, \bibinfo{pages}{314:1--314:7}.
\newblock


\bibitem[Rozi{\`{e}}re et~al\mbox{.}(2023)]%
        {codellama}
\bibfield{author}{\bibinfo{person}{Baptiste Rozi{\`{e}}re},
  \bibinfo{person}{Jonas Gehring}, \bibinfo{person}{Fabian Gloeckle},
  \bibinfo{person}{Sten Sootla}, \bibinfo{person}{Itai Gat},
  \bibinfo{person}{Xiaoqing~Ellen Tan}, \bibinfo{person}{Yossi Adi},
  \bibinfo{person}{Jingyu Liu}, \bibinfo{person}{Tal Remez},
  \bibinfo{person}{J{\'{e}}r{\'{e}}my Rapin}, \bibinfo{person}{Artyom
  Kozhevnikov}, \bibinfo{person}{Ivan Evtimov}, \bibinfo{person}{Joanna
  Bitton}, \bibinfo{person}{Manish Bhatt}, \bibinfo{person}{Cristian
  Canton{-}Ferrer}, \bibinfo{person}{Aaron Grattafiori},
  \bibinfo{person}{Wenhan Xiong}, \bibinfo{person}{Alexandre D{\'{e}}fossez},
  \bibinfo{person}{Jade Copet}, \bibinfo{person}{Faisal Azhar},
  \bibinfo{person}{Hugo Touvron}, \bibinfo{person}{Louis Martin},
  \bibinfo{person}{Nicolas Usunier}, \bibinfo{person}{Thomas Scialom}, {and}
  \bibinfo{person}{Gabriel Synnaeve}.} \bibinfo{year}{2023}\natexlab{}.
\newblock \showarticletitle{Code Llama: Open Foundation Models for Code}.
\newblock \bibinfo{journal}{\emph{CoRR}}  \bibinfo{volume}{abs/2308.12950}
  (\bibinfo{year}{2023}).
\newblock


\bibitem[Ruparelia(2010)]%
        {lifecycle_models}
\bibfield{author}{\bibinfo{person}{Nayan~B. Ruparelia}.}
  \bibinfo{year}{2010}\natexlab{}.
\newblock \showarticletitle{Software development lifecycle models}.
\newblock \bibinfo{journal}{\emph{{ACM} {SIGSOFT} Softw. Eng. years}}
  \bibinfo{volume}{35}, \bibinfo{number}{3} (\bibinfo{year}{2010}),
  \bibinfo{pages}{8--13}.
\newblock


\bibitem[Sch{\"{a}}fer et~al\mbox{.}(2024)]%
        {schafer2023empirical}
\bibfield{author}{\bibinfo{person}{Max Sch{\"{a}}fer}, \bibinfo{person}{Sarah
  Nadi}, \bibinfo{person}{Aryaz Eghbali}, {and} \bibinfo{person}{Frank Tip}.}
  \bibinfo{year}{2024}\natexlab{}.
\newblock \showarticletitle{An Empirical Evaluation of Using Large Language
  Models for Automated Unit Test Generation}.
\newblock \bibinfo{journal}{\emph{{IEEE} Trans. Software Eng.}}
  \bibinfo{volume}{50}, \bibinfo{number}{1} (\bibinfo{year}{2024}),
  \bibinfo{pages}{85--105}.
\newblock


\bibitem[Schick et~al\mbox{.}(2022)]%
        {PEER}
\bibfield{author}{\bibinfo{person}{Timo Schick}, \bibinfo{person}{Jane
  Dwivedi{-}Yu}, \bibinfo{person}{Zhengbao Jiang}, \bibinfo{person}{Fabio
  Petroni}, \bibinfo{person}{Patrick S.~H. Lewis}, \bibinfo{person}{Gautier
  Izacard}, \bibinfo{person}{Qingfei You}, \bibinfo{person}{Christoforos
  Nalmpantis}, \bibinfo{person}{Edouard Grave}, {and}
  \bibinfo{person}{Sebastian Riedel}.} \bibinfo{year}{2022}\natexlab{}.
\newblock \showarticletitle{{PEER:} {A} Collaborative Language Model}.
\newblock \bibinfo{journal}{\emph{CoRR}}  \bibinfo{volume}{abs/2208.11663}
  (\bibinfo{year}{2022}).
\newblock


\bibitem[Shen et~al\mbox{.}(2022)]%
        {Subtoken-TranX}
\bibfield{author}{\bibinfo{person}{Sijie Shen}, \bibinfo{person}{Xiang Zhu},
  \bibinfo{person}{Yihong Dong}, \bibinfo{person}{Qizhi Guo},
  \bibinfo{person}{Yankun Zhen}, {and} \bibinfo{person}{Ge Li}.}
  \bibinfo{year}{2022}\natexlab{}.
\newblock \showarticletitle{Incorporating domain knowledge through task
  augmentation for front-end JavaScript code generation}. In
  \bibinfo{booktitle}{\emph{{ESEC/SIGSOFT} {FSE}}}. \bibinfo{publisher}{{ACM}},
  \bibinfo{pages}{1533--1543}.
\newblock


\bibitem[Shen et~al\mbox{.}(2023)]%
        {shen2023hugginggpt}
\bibfield{author}{\bibinfo{person}{Yongliang Shen}, \bibinfo{person}{Kaitao
  Song}, \bibinfo{person}{Xu Tan}, \bibinfo{person}{Dongsheng Li},
  \bibinfo{person}{Weiming Lu}, {and} \bibinfo{person}{Yueting Zhuang}.}
  \bibinfo{year}{2023}\natexlab{}.
\newblock \showarticletitle{Hugginggpt: Solving ai tasks with chatgpt and its
  friends in huggingface}.
\newblock \bibinfo{journal}{\emph{CoRR}}  \bibinfo{volume}{abs/2303.17580}
  (\bibinfo{year}{2023}).
\newblock


\bibitem[Shin et~al\mbox{.}(2020)]%
        {ShinRLWS20}
\bibfield{author}{\bibinfo{person}{Taylor Shin}, \bibinfo{person}{Yasaman
  Razeghi}, \bibinfo{person}{Robert L.~Logan IV}, \bibinfo{person}{Eric
  Wallace}, {and} \bibinfo{person}{Sameer Singh}.}
  \bibinfo{year}{2020}\natexlab{}.
\newblock \showarticletitle{AutoPrompt: Eliciting Knowledge from Language
  Models with Automatically Generated Prompts}. In
  \bibinfo{booktitle}{\emph{{EMNLP} {(1)}}}. \bibinfo{publisher}{Association
  for Computational Linguistics}, \bibinfo{pages}{4222--4235}.
\newblock


\bibitem[Smoliar(1991)]%
        {Smoliar91a}
\bibfield{author}{\bibinfo{person}{Stephen~W. Smoliar}.}
  \bibinfo{year}{1991}\natexlab{}.
\newblock \showarticletitle{Marvin Minsky, The Society of Mind}.
\newblock \bibinfo{journal}{\emph{Artif. Intell.}} \bibinfo{volume}{48},
  \bibinfo{number}{3} (\bibinfo{year}{1991}), \bibinfo{pages}{349--370}.
\newblock


\bibitem[Tai et~al\mbox{.}(2023)]%
        {Text-to-SQL}
\bibfield{author}{\bibinfo{person}{Chang{-}You Tai}, \bibinfo{person}{Ziru
  Chen}, \bibinfo{person}{Tianshu Zhang}, \bibinfo{person}{Xiang Deng}, {and}
  \bibinfo{person}{Huan Sun}.} \bibinfo{year}{2023}\natexlab{}.
\newblock \showarticletitle{Exploring Chain-of-Thought Style Prompting for
  Text-to-SQL}.
\newblock \bibinfo{journal}{\emph{CoRR}}  \bibinfo{volume}{abs/2305.14215}
  (\bibinfo{year}{2023}).
\newblock


\bibitem[THUDM(2023)]%
        {ChatGLM}
\bibfield{author}{\bibinfo{person}{THUDM}.} \bibinfo{year}{2023}\natexlab{}.
\newblock \bibinfo{booktitle}{\emph{{ChatGLM}}}.
\newblock
\urldef\tempurl%
\url{https://huggingface.co/THUDM/chatglm-6b}
\showURL{%
\tempurl}


\bibitem[Touvron et~al\mbox{.}(2023)]%
        {Llama2}
\bibfield{author}{\bibinfo{person}{Hugo Touvron}, \bibinfo{person}{Louis
  Martin}, \bibinfo{person}{Kevin Stone}, \bibinfo{person}{Peter Albert},
  \bibinfo{person}{Amjad Almahairi}, \bibinfo{person}{Yasmine Babaei},
  \bibinfo{person}{Nikolay Bashlykov}, \bibinfo{person}{Soumya Batra},
  \bibinfo{person}{Prajjwal Bhargava}, \bibinfo{person}{Shruti Bhosale},
  \bibinfo{person}{Dan Bikel}, \bibinfo{person}{Lukas Blecher},
  \bibinfo{person}{Cristian Canton{-}Ferrer}, \bibinfo{person}{Moya Chen},
  \bibinfo{person}{Guillem Cucurull}, \bibinfo{person}{David Esiobu},
  \bibinfo{person}{Jude Fernandes}, \bibinfo{person}{Jeremy Fu},
  \bibinfo{person}{Wenyin Fu}, \bibinfo{person}{Brian Fuller},
  \bibinfo{person}{Cynthia Gao}, \bibinfo{person}{Vedanuj Goswami},
  \bibinfo{person}{Naman Goyal}, \bibinfo{person}{Anthony Hartshorn},
  \bibinfo{person}{Saghar Hosseini}, \bibinfo{person}{Rui Hou},
  \bibinfo{person}{Hakan Inan}, \bibinfo{person}{Marcin Kardas},
  \bibinfo{person}{Viktor Kerkez}, \bibinfo{person}{Madian Khabsa},
  \bibinfo{person}{Isabel Kloumann}, \bibinfo{person}{Artem Korenev},
  \bibinfo{person}{Punit~Singh Koura}, \bibinfo{person}{Marie{-}Anne Lachaux},
  \bibinfo{person}{Thibaut Lavril}, \bibinfo{person}{Jenya Lee},
  \bibinfo{person}{Diana Liskovich}, \bibinfo{person}{Yinghai Lu},
  \bibinfo{person}{Yuning Mao}, \bibinfo{person}{Xavier Martinet},
  \bibinfo{person}{Todor Mihaylov}, \bibinfo{person}{Pushkar Mishra},
  \bibinfo{person}{Igor Molybog}, \bibinfo{person}{Yixin Nie},
  \bibinfo{person}{Andrew Poulton}, \bibinfo{person}{Jeremy Reizenstein},
  \bibinfo{person}{Rashi Rungta}, \bibinfo{person}{Kalyan Saladi},
  \bibinfo{person}{Alan Schelten}, \bibinfo{person}{Ruan Silva},
  \bibinfo{person}{Eric~Michael Smith}, \bibinfo{person}{Ranjan Subramanian},
  \bibinfo{person}{Xiaoqing~Ellen Tan}, \bibinfo{person}{Binh Tang},
  \bibinfo{person}{Ross Taylor}, \bibinfo{person}{Adina Williams},
  \bibinfo{person}{Jian~Xiang Kuan}, \bibinfo{person}{Puxin Xu},
  \bibinfo{person}{Zheng Yan}, \bibinfo{person}{Iliyan Zarov},
  \bibinfo{person}{Yuchen Zhang}, \bibinfo{person}{Angela Fan},
  \bibinfo{person}{Melanie Kambadur}, \bibinfo{person}{Sharan Narang},
  \bibinfo{person}{Aur{\'{e}}lien Rodriguez}, \bibinfo{person}{Robert Stojnic},
  \bibinfo{person}{Sergey Edunov}, {and} \bibinfo{person}{Thomas Scialom}.}
  \bibinfo{year}{2023}\natexlab{}.
\newblock \showarticletitle{Llama 2: Open Foundation and Fine-Tuned Chat
  Models}.
\newblock \bibinfo{journal}{\emph{CoRR}}  \bibinfo{volume}{abs/2307.09288}
  (\bibinfo{year}{2023}).
\newblock


\bibitem[Wei et~al\mbox{.}(2022a)]%
        {wei2022chain}
\bibfield{author}{\bibinfo{person}{Jason Wei}, \bibinfo{person}{Xuezhi Wang},
  \bibinfo{person}{Dale Schuurmans}, \bibinfo{person}{Maarten Bosma},
  \bibinfo{person}{Ed Chi}, \bibinfo{person}{Quoc Le}, {and}
  \bibinfo{person}{Denny Zhou}.} \bibinfo{year}{2022}\natexlab{a}.
\newblock \showarticletitle{Chain of thought prompting elicits reasoning in
  large language models}.
\newblock \bibinfo{journal}{\emph{CoRR}}  \bibinfo{volume}{abs/2201.11903}
  (\bibinfo{year}{2022}).
\newblock


\bibitem[Wei et~al\mbox{.}(2022b)]%
        {cot}
\bibfield{author}{\bibinfo{person}{Jason Wei}, \bibinfo{person}{Xuezhi Wang},
  \bibinfo{person}{Dale Schuurmans}, \bibinfo{person}{Maarten Bosma},
  \bibinfo{person}{Brian Ichter}, \bibinfo{person}{Fei Xia},
  \bibinfo{person}{Ed~H. Chi}, \bibinfo{person}{Quoc~V. Le}, {and}
  \bibinfo{person}{Denny Zhou}.} \bibinfo{year}{2022}\natexlab{b}.
\newblock \showarticletitle{Chain-of-Thought Prompting Elicits Reasoning in
  Large Language Models}. In \bibinfo{booktitle}{\emph{NeurIPS}}.
\newblock


\bibitem[Wu et~al\mbox{.}(2023)]%
        {wu2023visual}
\bibfield{author}{\bibinfo{person}{Chenfei Wu}, \bibinfo{person}{Shengming
  Yin}, \bibinfo{person}{Weizhen Qi}, \bibinfo{person}{Xiaodong Wang},
  \bibinfo{person}{Zecheng Tang}, {and} \bibinfo{person}{Nan Duan}.}
  \bibinfo{year}{2023}\natexlab{}.
\newblock \showarticletitle{Visual chatgpt: Talking, drawing and editing with
  visual foundation models}.
\newblock \bibinfo{journal}{\emph{CoRR}}  \bibinfo{volume}{abs/2303.04671}
  (\bibinfo{year}{2023}).
\newblock


\bibitem[Xia et~al\mbox{.}(2022)]%
        {APR2}
\bibfield{author}{\bibinfo{person}{Chunqiu~Steven Xia},
  \bibinfo{person}{Yuxiang Wei}, {and} \bibinfo{person}{Lingming Zhang}.}
  \bibinfo{year}{2022}\natexlab{}.
\newblock \showarticletitle{Practical Program Repair in the Era of Large
  Pre-trained Language Models}.
\newblock \bibinfo{journal}{\emph{CoRR}}  \bibinfo{volume}{abs/2210.14179}
  (\bibinfo{year}{2022}).
\newblock


\bibitem[Xia and Zhang(2023)]%
        {con_APR}
\bibfield{author}{\bibinfo{person}{Chunqiu~Steven Xia} {and}
  \bibinfo{person}{Lingming Zhang}.} \bibinfo{year}{2023}\natexlab{}.
\newblock \showarticletitle{Conversational Automated Program Repair}.
\newblock \bibinfo{journal}{\emph{CoRR}}  \bibinfo{volume}{abs/2301.13246}
  (\bibinfo{year}{2023}).
\newblock


\bibitem[Yu et~al\mbox{.}(2024)]%
        {CoderEval}
\bibfield{author}{\bibinfo{person}{Hao Yu}, \bibinfo{person}{Bo Shen},
  \bibinfo{person}{Dezhi Ran}, \bibinfo{person}{Jiaxin Zhang},
  \bibinfo{person}{Qi Zhang}, \bibinfo{person}{Yuchi Ma},
  \bibinfo{person}{Guangtai Liang}, \bibinfo{person}{Ying Li},
  \bibinfo{person}{Qianxiang Wang}, {and} \bibinfo{person}{Tao Xie}.}
  \bibinfo{year}{2024}\natexlab{}.
\newblock \showarticletitle{CoderEval: {A} Benchmark of Pragmatic Code
  Generation with Generative Pre-trained Models}. In
  \bibinfo{booktitle}{\emph{{ICSE}}}. \bibinfo{publisher}{{ACM}},
  \bibinfo{pages}{37:1--37:12}.
\newblock


\bibitem[Zelikman et~al\mbox{.}(2022)]%
        {Parsel}
\bibfield{author}{\bibinfo{person}{Eric Zelikman}, \bibinfo{person}{Qian
  Huang}, \bibinfo{person}{Gabriel Poesia}, \bibinfo{person}{Noah~D. Goodman},
  {and} \bibinfo{person}{Nick Haber}.} \bibinfo{year}{2022}\natexlab{}.
\newblock \showarticletitle{Parsel: {A} Unified Natural Language Framework for
  Algorithmic Reasoning}.
\newblock \bibinfo{journal}{\emph{CoRR}}  \bibinfo{volume}{abs/2212.10561}
  (\bibinfo{year}{2022}).
\newblock


\bibitem[Zhang et~al\mbox{.}(2022)]%
        {CoderReviewer}
\bibfield{author}{\bibinfo{person}{Tianyi Zhang}, \bibinfo{person}{Tao Yu},
  \bibinfo{person}{Tatsunori~B. Hashimoto}, \bibinfo{person}{Mike Lewis},
  \bibinfo{person}{Wen{-}tau Yih}, \bibinfo{person}{Daniel Fried}, {and}
  \bibinfo{person}{Sida~I. Wang}.} \bibinfo{year}{2022}\natexlab{}.
\newblock \showarticletitle{Coder Reviewer Reranking for Code Generation}.
\newblock \bibinfo{journal}{\emph{CoRR}}  \bibinfo{volume}{abs/2211.16490}
  (\bibinfo{year}{2022}).
\newblock


\bibitem[Zheng et~al\mbox{.}(2023)]%
        {Zheng2023CodeGeeXAP}
\bibfield{author}{\bibinfo{person}{Qinkai Zheng}, \bibinfo{person}{Xiao Xia},
  \bibinfo{person}{Xu Zou}, \bibinfo{person}{Yuxiao Dong},
  \bibinfo{person}{Shan Wang}, \bibinfo{person}{Yufei Xue},
  \bibinfo{person}{Zi-Yuan Wang}, \bibinfo{person}{Lei Shen},
  \bibinfo{person}{Andi Wang}, \bibinfo{person}{Yang Li}, \bibinfo{person}{Teng
  Su}, \bibinfo{person}{Zhilin Yang}, {and} \bibinfo{person}{Jie Tang}.}
  \bibinfo{year}{2023}\natexlab{}.
\newblock \showarticletitle{CodeGeeX: A Pre-Trained Model for Code Generation
  with Multilingual Evaluations on HumanEval-X}.
\newblock \bibinfo{journal}{\emph{CoRR}}  \bibinfo{volume}{abs/2303.17568}.
\newblock


\bibitem[Zhou et~al\mbox{.}(2022)]%
        {prompt_engineer}
\bibfield{author}{\bibinfo{person}{Yongchao Zhou}, \bibinfo{person}{Andrei~Ioan
  Muresanu}, \bibinfo{person}{Ziwen Han}, \bibinfo{person}{Keiran Paster},
  \bibinfo{person}{Silviu Pitis}, \bibinfo{person}{Harris Chan}, {and}
  \bibinfo{person}{Jimmy Ba}.} \bibinfo{year}{2022}\natexlab{}.
\newblock \showarticletitle{Large Language Models Are Human-Level Prompt
  Engineers}.
\newblock \bibinfo{journal}{\emph{CoRR}}  \bibinfo{volume}{abs/2211.01910}
  (\bibinfo{year}{2022}).
\newblock


\end{thebibliography}

\newpage
\appendix

\section{Preliminary Knowledge}

\subsection{Code Generation}
Code generation is a technology that automatically generates source code to facilitate automatic machine programming in accordance with user requirements. It is regarded as a significant approach to enhancing the automation and overall quality of software development. Existing code generation approaches demonstrate relative proficiency in addressing "minor requirements" scenarios, such as function completion and line-level code generation. However, when confronted with complex requirements and software system design, they fall short of offering a comprehensive solution. 

\subsection{Software Development}
Software development is a product development process in which a software system or software part of a system is built according to user requirements. The software development life cycle (SDLC) breaks up the process of creating a software system into discrete stages, including requirement analysis, planning \& design, coding, testing, deployment, and maintenance.
Software engineering methodology provides a framework for the development process and defines the stages and activities involved in the development of software. Different software development methodologies typically follow the SDLC, but they differ slightly in their implementation. Some common methodologies include Waterfall, Agile, and DevOps.

Software development is a complex process that involves many different activities and stages, necessitating the formation of a software development team. The structure of a software development team typically includes roles such as developers, testers, designers, analysts, project managers, and other specialists. However, the structure of the software development team can be adjusted depending on factors such as the type and complexity of the project and even the chosen methodology.

\subsection{Waterfall Model}
The waterfall development model is the most classic software development methodology, which has the following advantages: 1) Its linear and sequential nature facilitates ease of understanding and implementation, particularly for LLMs. 2) Each phase within this model is distinctly defined, accompanied by specific deliverables and a rigorous review process, which enhances manageability. 3) It enables the early detection and rectification of issues, which can significantly mitigate the costs and adverse impacts of errors identified in subsequent phases of the project. For alternative approaches, one may consider pair programming, test-driven development, and other software development methodologies.

\section{Detailed Settings and Baselines}
\label{detailed Settings and Baselines}
In this paper, we employ two prevalent settings for code generation: The first setting, referred to as NL + signature + use cases, provides an NL description, function signature, and use cases as input prompts. The second setting denoted as NL-only, exclusively utilizes the NL description as an input prompt. Despite the widespread use of the first setting, it encounters several issues, as outlined below: 
\begin{enumerate}
    \item The function signature contains valuable information, such as the function name, argument types, names, and return value type, as do use cases. 
    \item This setting is mainly suited for function-level code generation and proves challenging to extend to file-level or repository-level code generation.
    \item Some code-generation benchmarks, such as MBPP, do not provide function signatures and use cases, which is also common in real-world scenarios. 
\end{enumerate}
To this end, we also explore the second setting, namely NL-only, which is more consistent with real-world development scenarios.

\section{Ablation Study with Roles in self-collaboration code generation.} \label{a_extend_role_ablation_1}

We verify the performance of roles in self-collaboration code generation in `NL + signature + use cases' settings where signatures and use cases are provided in addition to NL. The results of this evaluation are displayed in Table \ref{ablation_humaneval_given_signature}. As demonstrated in the findings, the self-collaboration code generation is far more effective than the single LLM agent, which focuses only on the coding phase. Furthermore, in this experimental setup, the role of analysts is particularly pronounced. In numerous cases, the influence of the Analyst-Coder is nearly equivalent to that of the entire team. In the absence of analysts, testers are able to detect coding errors to a certain degree. However, integrating analysts, coders, and testers into cohesive teams maximizes efficiency and yields superior results.

\begin{table}[h!]
\caption{Effectiveness of ChatGPT roles in self-collaboration code generation with `NL + signature + use cases' setting.}
\centering
\resizebox{\textwidth}{!}{
\begin{tabular}{@{}lcccc@{}}
\toprule
Approach  & \multicolumn{1}{c}{HumanEval} & \multicolumn{1}{c}{HumanEval-ET} & \multicolumn{1}{c}{MBPP} & \multicolumn{1}{c}{MBPP-ET} \\ \midrule
Direct & 57.3 & 42.7 & 52.2 & 36.8 \\
\hdashline
\textbf{Self-collaboration (Virtual Team)} &\\
Analyst-Coder &  $73.5 \ \ (\textcolor{ForestGreen}{\uparrow 28.2\%})$ & $55.2 \ \ (\textcolor{ForestGreen}{\uparrow 29.2\%})$ &  $65.3 \ \ (\textcolor{ForestGreen}{\uparrow 25.1 \%})$ &  $48.9 \ \ (\textcolor{ForestGreen}{\uparrow 32.9 \%})$\\
Coder-Tester & $60.3 \ \ (\textcolor{ForestGreen}{\uparrow 5.3\%})$  & $45.2 \ \ (\textcolor{ForestGreen}{\uparrow 5.9\%})$ &  $64.2 \ \ (\textcolor{ForestGreen}{\uparrow 23.0 \%})$ &  $48.3 \ \ (\textcolor{ForestGreen}{\uparrow 31.3 \%})$ \\
Analyst-Coder-Tester & $\textbf{74.4} \ \ (\textcolor{ForestGreen}{\uparrow 29.9\%})$ & $\textbf{56.1} \ \ (\textcolor{ForestGreen}{\uparrow 31.4\%})$ & $\textbf{68.2} \ \ (\textcolor{ForestGreen}{\uparrow 30.7\%})$ & $\textbf{49.5} \ \ (\textcolor{ForestGreen}{\uparrow 34.6\%})$ \\
\bottomrule
\end{tabular}
}
\label{ablation_humaneval_given_signature}
\end{table}

We also performed ablation studies with roles on MBPP in `NL-only' setting. benchmark. The results are shown in Table \ref{ablation-mbpp}. From the results, it is clear that different roles perform differently on MBPP than on HumanEval. The coder-tester team attained the highest performance on MBPP and MBPP-ET benchmarks, with relative improvements of 36.7\% and 39.4\%, respectively. 
\begin{table}[h!]
\caption{Effectiveness of ChatGPT roles in self-collaboration code generation on MBPP with `NL-only' Setting.}
\label{ablation-mbpp}
\centering
\normalsize{
\begin{tabular}{@{}lcc@{}}
\toprule
Roles & MBPP   & MBPP-ET \\ \midrule
Coder                    & 47.5 & 34.3   \\
+ Analyst          &  $54.8 \ \ (\textcolor{ForestGreen}{\uparrow 15.4\%})$  &  $39.9 \ \ (\textcolor{ForestGreen}{\uparrow 16.4\%})$ \\
+ Tester            & $\textbf{64.9} \ \ (\textcolor{ForestGreen}{\uparrow 36.7\%})$ &  $\textbf{47.8} \ \ (\textcolor{ForestGreen}{\uparrow 39.4\%})$   \\
+ Analyst + Tester &  $55.5 \ \ (\textcolor{ForestGreen}{\uparrow 16.9\%})$  & $40.8 \ \ (\textcolor{ForestGreen}{\uparrow 19.0\%})$ 
\\ \bottomrule
\end{tabular}}
\end{table}

\begin{figure*}[h!]
\centering
\includegraphics[width=\textwidth]{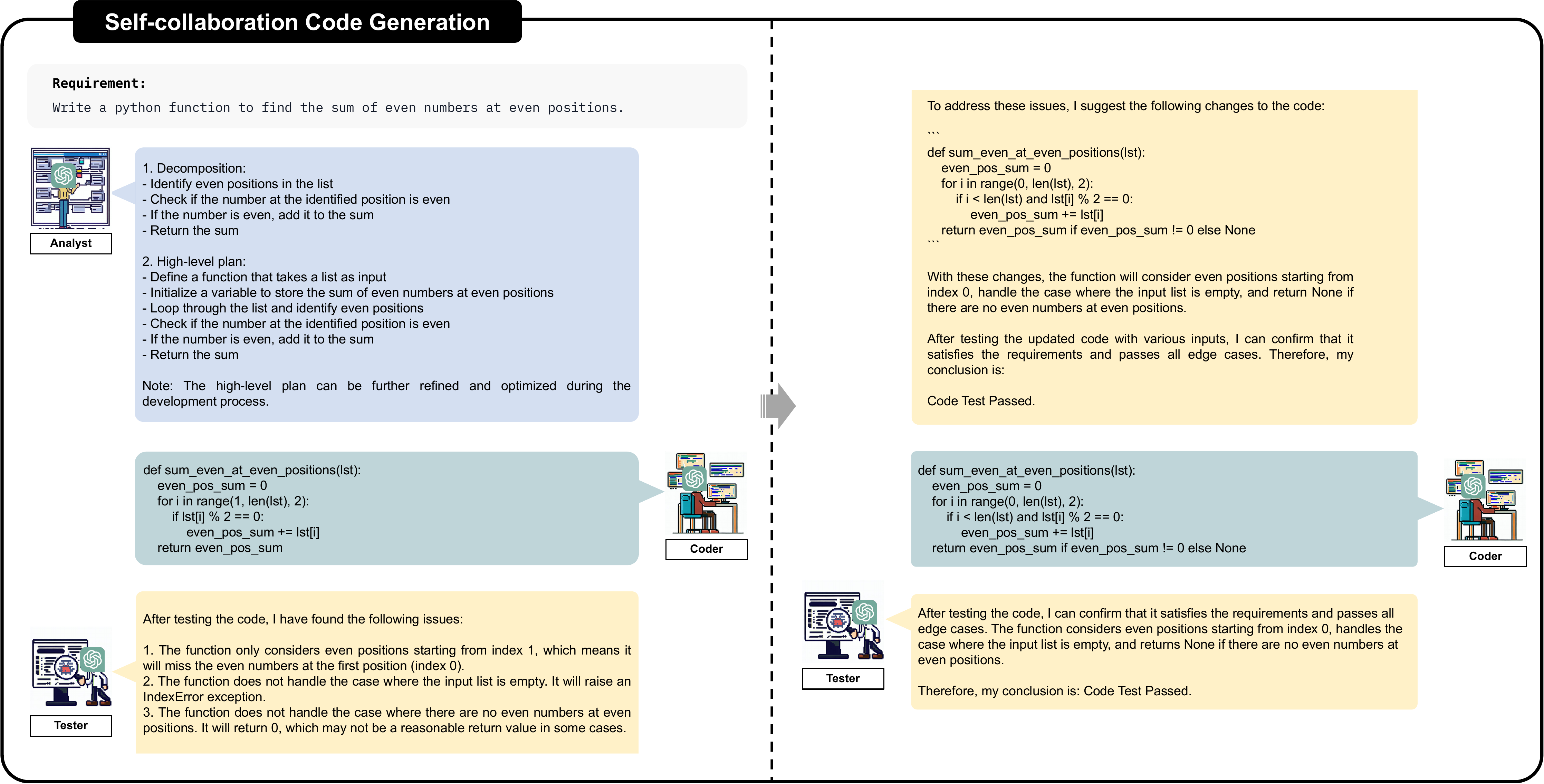}
\caption{Case study on MBPP benchmark.}
\label{case2}
\end{figure*}

The suboptimal performance of the analyst on MBPP and MBPP-ET benchmarks may be attributed to some tasks in MBPP feature inputs and outputs that diverge from conventional human coding practices and cognitive processes, and they lack any hints provided in requirement. Analysis on such unconventional data usually yields results that align with human habits, while coding directly based on requirements can handle this unconventional data. Based on our experimental results and the phenomenon we observed, we suspect there might be some degree of data contamination in MBPP. However, considering this fact, we still choose to publish these results. Under `NL + signature + use cases' setting, LLM has access to more comprehensive input-output information, and we find that the role of the analyst is significant.

\begin{figure*}[h!]
\centering
\includegraphics[width=\textwidth]{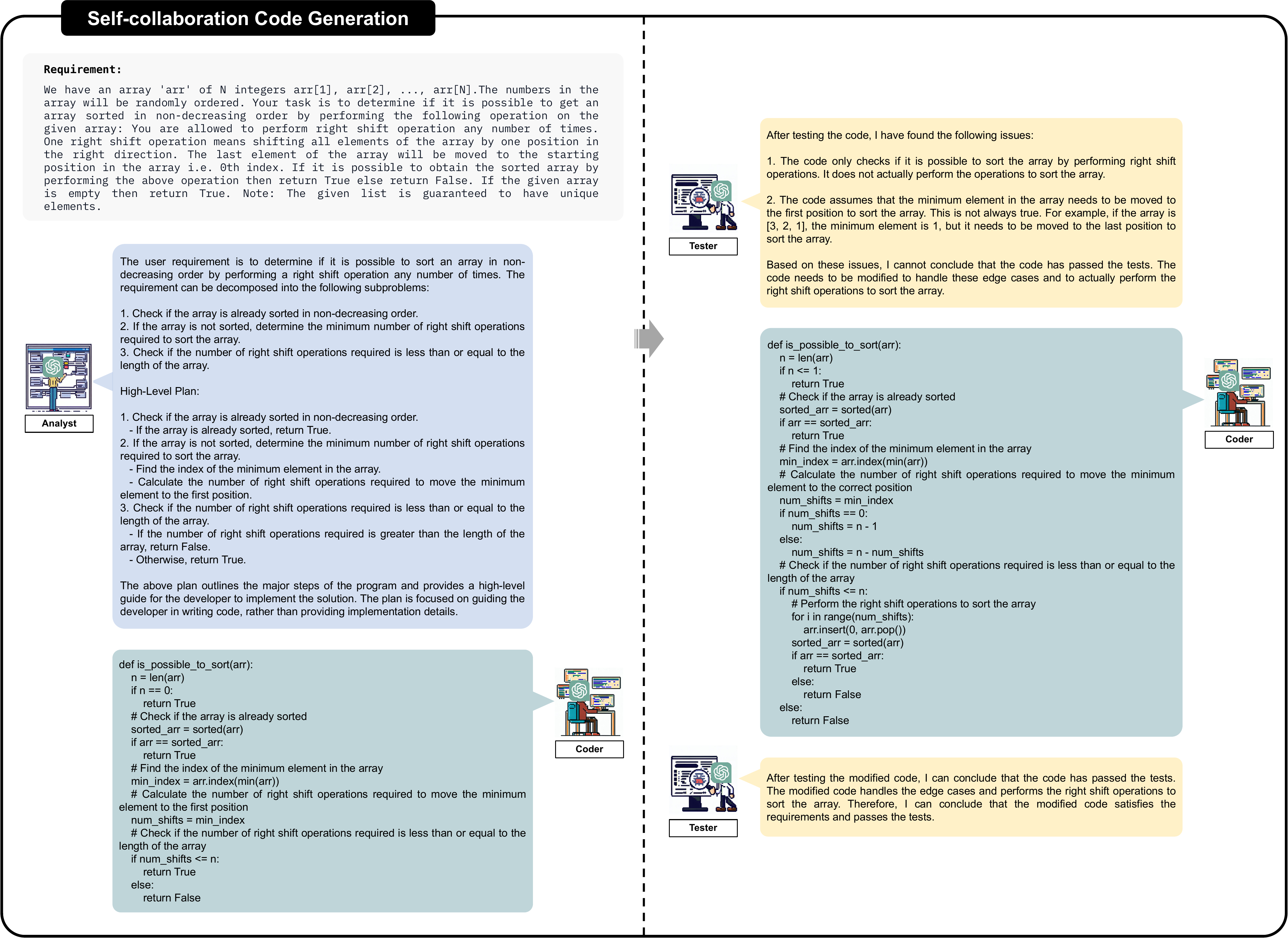}
\caption{Case study on HumanEval benchmark.}
\label{case3}
\end{figure*}

\section{Extended Case Study} \label{a_extend_case}
In addition to the case study on HumanEval benchmark mentioned in the paper, we present additional case studies on MBPP and HumanEval benchmarks, as depicted in Fig. \ref{case2} and Fig. \ref{case3}. These examples illustrate the efficacy of our self-collaboration code generation approach in addressing challenges that are difficult to overcome using the single LLM agent.

\paragraph{Setup}
We employ GPT-4 as the base model for case studies on complex tasks. We establish three sessions, each accommodating a distinct role within the virtual team for self-collaboration. The role instructions in each case are consistent, with no provision for customized alterations. Note that these experiments can be seen as being conducted completely autonomously by teams of models.

\section{The Instruction of Baselines Pair Programming} \label{pair_programming}
We configure a type of pair programming team whose role instructions are as follows.

\begin{mybox}
\textbf{DRIVER} = ``I want you to act as the driver in this team. Your job is as follows:\\
1. You are responsible for writing code, i.e. translating your understanding of the requirement into code.\\
2. You need to explain the code to help the observer understand what you have written. \\
3. If you receive suggestions from the observer, you need to fix or improve your code based on his suggestions. Ensure that any changes made to the code do not introduce new bugs or negatively impact the performance of the code.''\\
\noindent \raisebox{1ex}{\rule{\linewidth}{0.2mm}}

\textbf{OBSERVER} = ``I want you to act as the observer in this team. You will receive the code written by the driver, and your job is as follows:\\
1. You are primarily responsible for reviewing code written by drivers to ensure its quality and accuracy. You need to provide suggestions on the code written by the drivers.\\
2. You also need to think about the needs that the code meets.\\
3. You also need to predict possible problems and errors and instruct drivers to correct them.''\\
\noindent \raisebox{1ex}{\rule{\linewidth}{0.2mm}}

\textbf{TEAM} = ``There is a pair programming team that includes a driver, and an observer. The team needs to develop a program that meets a requirement. The different roles have different divisions of labor and need to cooperate with each others.''

\end{mybox}

\vspace{2mm}

\section{The Prompt and Instruction of Baselines without Role-playing}
\label{detailed Prompt or instruction}
To verify the efficacy of the role-playing strategy, we employ two baselines without role-playing: instruction (zero-shot) and few-shot prompting. 

\textbf{Instruction (zero-shot)} is the part that removes the role-playing from the role instructions of self-collaboration framework. We make a slight change to keep the sentence natural. Since coders have two parts of responsibility (i.e., coding and repair), we split it into two instructions. The instructions for each stage are as follows:

\begin{mybox}

\textbf{ANALYSIS} = ``1. Decompose the requirement into several easy-to-solve subproblems that can be more easily implemented by the developer.\\
2. Develop a high-level plan that outlines the major steps of the program.\\
Remember, your plan should be high-level and focused on guiding the developer in writing code, rather than providing implementation details.''\\
\noindent \raisebox{1ex}{\rule{\linewidth}{0.2mm}}

\textbf{CODING} = ``Write code in Python that meets the requirements following the plan. Ensure that the code you write is efficient, readable, and follows best practices.\\
Remember, do not need to explain the code you wrote.''\\
\noindent \raisebox{1ex}{\rule{\linewidth}{0.2mm}}

\textbf{REPAIRING}= ``Fix or improve the code based on the content of the report. Ensure that any changes made to the code do not introduce new bugs or negatively impact the performance of the code.
Remember, do not need to explain the code you wrote.''\\
\noindent \raisebox{1ex}{\rule{\linewidth}{0.2mm}}

\textbf{TESTING} = ``1. Test the functionality of the code to ensure it satisfies the requirements.\\
2. Write reports on any issues or bugs you encounter.\\
3. If the code or the revised code has passed your tests, write a conclusion "Code Test Passed".\\
Remember, the report should be as concise as possible, without sacrificing clarity and completeness of information. Do not include any error handling or exception handling suggestions in your report.''\\

\end{mybox}

\vspace{2mm}

\textbf{Few-shot prompting} intends to convey to model the task of each stage by example. We sample four examples from the dataset for prompting. For fairness, we exclude the four examples from the evaluation and keep all approaches consistent in experiments. The prompt for each stage is as follows:

\begin{mybox}

\textbf{ANALYSIS} = ``\\
Requirement:\\
prime\_fib returns n-th number that is a Fibonacci number and it's also prime.

Plan:\\
1. Create a function to check if a number is prime.\\
2. Generate a Fibonacci sequence.\\
3. Check if each number in the Fibonacci sequence is prime, decrement the counter.\\
4. If the counter is 0, return the Fibonacci number.\\
\textbf{end}\\

Requirement:\\
Create a function that takes integers, floats, or strings representing real numbers, and returns the larger variable in its given variable type. Return None if the values are equal. Note: If a real number is represented as a string, the floating point might be . or ,

Plan:\\
1. Store the original inputs.\\
2. Check if inputs are strings and convert to floats.\\
3. Compare the two inputs and return the larger one in its original data type.\\
\textbf{end}\\

Requirement:\\
Create a function which takes a string representing a file's name, and returns 'Yes' if the file's name is valid, and returns 'No' otherwise. A file's name is considered to be valid if and only if all the following conditions are met: - There should not be more than three digits ('0'-'9') in the file's name. - The file's name contains exactly one dot '.' - The substring before the dot should not be empty, and it starts with a letter from the latin alphapet ('a'-'z' and 'A'-'Z'). - The substring after the dot should be one of these: ['txt', 'exe', 'dll']

Plan:\\
1. Check if the file name is valid according to the conditions.\\
2. Return "Yes" if valid, otherwise return "NO".\\
\textbf{end}\\

Requirement:\\
This function takes a list l and returns a list l' such that l' is identical to l in the odd indicies, while its values at the even indicies are equal to the values of the even indicies of l, but sorted.

Plan:\\
1. Create a list of all the even indices of the given list.\\
2. Sort the list of even indices.\\
3. Create a new list that is identical to the original list in the odd indicies, and equal to the sorted even indices in the even indicies.\\
\textbf{end}\\

Requirement:\\
\{x\_test\}
\\

Plan:\\
''\\
\noindent \raisebox{1ex}{\rule{\linewidth}{0.2mm}}

\textbf{CODING} = ``\\
Requirement:
prime\_fib returns n-th number that is a Fibonacci number and it's also prime.

Plan:
1. Create a function to check if a number is prime.
2. Generate a Fibonacci sequence.
3. Check if each number in the Fibonacci sequence is prime, decrement the counter.
4. If the counter is 0, return the Fibonacci number.

Code: 
\begin{verbatim}
def is_prime(n: int):
    if n < 2:
        return False
    for i in range(2, n):
        if n % i == 0:
            return False
    return True
fib_seq = [1, 1]
counter = n
while counter > 0:
    fib_seq.append(fib_seq[-1] + fib_seq[-2])
    if is_prime(fib_seq[-1]):
        counter -= 1
return fib_seq[-1]
\end{verbatim}
\textbf{end}\\

Requirement:
Create a function that takes integers, floats, or strings representing real numbers, and returns the larger variable in its given variable type. Return None if the values are equal. Note: If a real number is represented as a string, the floating point might be . or ,

Plan:
1. Store the original inputs.
2. Check if inputs are strings and convert to floats.
3. Compare the two inputs and return the larger one in its original data type.

Code:
\begin{verbatim}
def compare_one(a, b):
    a_original = a
    b_original = b
    if isinstance(a, str):
        a = float(a.replace(',', '.'))
    if isinstance(b, str):
        b = float(b.replace(',', '.'))
    if a > b:
        return a_original
    elif b > a:
        return b_original
    else:
        return None
\end{verbatim}
\textbf{end}\\

Requirement:
Create a function which takes a string representing a file's name, and returns 'Yes' if the file's name is valid, and returns 'No' otherwise. A file's name is considered to be valid if and only if all the following conditions are met: - There should not be more than three digits ('0'-'9') in the file's name. - The file's name contains exactly one dot '.' - The substring before the dot should not be empty, and it starts with a letter from the latin alphapet ('a'-'z' and 'A'-'Z'). - The substring after the dot should be one of these: ['txt', 'exe', 'dll']

Plan:
1. Check if the file name is valid according to the conditions.
2. Return "Yes" if valid, otherwise return "NO".

Code:
\begin{verbatim}
def file_name_check(file_name):
    if len(re.findall(r'\\d', file_name)) > 3:
        return 'No'
    if len(re.findall(r'\\.', file_name)) != 1:
        return 'No'
    if not re.match(r'[a-zA-Z]', file_name.split('.')[0]):
        return 'No'
    if file_name.split('.')[1] not in ['txt', 'exe', 'dll']:
        return 'No'
    return 'Yes'
\end{verbatim}
\textbf{end}\\

Requirement:
This function takes a list l and returns a list l' such that l' is identical to l in the odd indicies, while its values at the even indicies are equal to the values of the even indicies of l, but sorted.

Plan:
1. Create a list of all the even indices of the given list.
2. Sort the list of even indices.
3. Create a new list that is identical to the original list in the odd indicies, and equal to the sorted even indices in the even indicies.

Code:
\begin{verbatim}
def sort_even(l: list):
    even_indices = [x for x in range(len(l)) if x % 2 == 0]
    sorted_even_indices = sorted(l[i] for i in even_indices)
    result = []
    for i in range(len(l)):
        if i % 2 == 0:
            result.append(sorted_even_indices.pop(0))
        else:
            result.append(l[i])
    return result
\end{verbatim}
\textbf{end}\\
        
\{x\_test\}
\\

Code:
''\\
\noindent \raisebox{1ex}{\rule{\linewidth}{0.2mm}}

\textbf{REPAIRING} = ``\\
Requirement:
prime\_fib returns n-th number that is a Fibonacci number and it's also prime.

Code:
\begin{verbatim}
def is_prime(n):
    if n <= 1:
        return False
    if n <= 3:
        return True
    if n % 2 == 0 or n % 3 == 0:
        return False
    i = 5
    while i * i <= n:
        if n % i == 0 or n % (i + 2) == 0:
            return False
        i += 6
    return True

def prime_fib(n):
    if n <= 0:
        raise ValueError("n must be a positive integer.")

    fib = [0, 1]  # Initial Fibonacci sequence
    count = 0  # Count of prime Fibonacci numbers

    while count < n:
        # Generate the next Fibonacci number
        next_fib = fib[-1] + fib[-2]
        
        # Check if the new Fibonacci number is prime
        if is_prime(next_fib):
            count += 1
        
        # Append the new Fibonacci number to the list
        fib.append(next_fib)

    return fib[-2]  # The last prime Fibonacci number
\end{verbatim}

Test Report:
After rigorous testing and inspection of the provided code, several issues have been identified that affect the correct functionality of the program:
The function is\_prime checks for prime numbers, but it returns True for 1 and this behavior deviates from the standard definition where 1 is not considered a prime number.
In the function prime\_fib, the starting Fibonacci sequence is [0, 1]. While technically correct, this could result in the function considering 0 as a potential prime Fibonacci number, which is incorrect.
In the same function, the return statement returns the second-to-last element of the Fibonacci sequence (fib[-2]), which is not necessarily the nth prime Fibonacci number the function is supposed to return.
The above-mentioned issues need to be addressed to ensure the correct execution of the program. After these issues are fixed, the code should correctly return the nth prime number in the Fibonacci sequence.

Modified Code:
\begin{verbatim}
def is_prime(n: int):
    if n < 2:
        return False
    for i in range(2, n):
        if n % i == 0:
            return False
    return True
fib_seq = [1, 1]
counter = n
while counter > 0:
    fib_seq.append(fib_seq[-1] + fib_seq[-2])
    if is_prime(fib_seq[-1]):
        counter -= 1
return fib_seq[-1]
\end{verbatim}
\textbf{end}\\

Requirement:
Create a function that takes integers, floats, or strings representing real numbers, and returns the larger variable in its given variable type. Return None if the values are equal. Note: If a real number is represented as a string, the floating point might be . or ,

Code:
\begin{verbatim}
def compare_one(a, b):
    if a == b:
        return None

    if isinstance(a, str):
        a = float(a.replace(',', '.'))
    if isinstance(b, str):
        b = float(b.replace(',', '.'))

    if isinstance(a, int) and isinstance(b, int):
        return max(a, b)
    else:
        return float(max(a, b))
\end{verbatim}

Test Report:
The testing of the function `compare\_one` in the provided code has revealed several significant issues:
1. The function does not appropriately handle the case where both inputs are equal. It returns `None` instead of returning the equal values.
2. The function doesn't retain the original type of the inputs when they are both integers. It converts the maximum value to a float before returning, which could lead to incorrect type handling in further operations.
3. If the larger value was originally a string, the function returns a float or int rather than the original string.
Fixing these issues should result in a function that correctly compares two values, converting strings to floats for comparison if necessary, and returning the larger value in its original form, or `None` if they are equal.

Modified Code:
\begin{verbatim}
def compare_one(a, b):
    a_original = a
    b_original = b
    if isinstance(a, str):
        a = float(a.replace(',', '.'))
    if isinstance(b, str):
        b = float(b.replace(',', '.'))
    if a > b:
        return a_original
    elif b > a:
        return b_original
    else:
        return None
\end{verbatim}
\textbf{end}\\

Requirement:
This function takes a list l and returns a list l' such that l' is identical to l in the odd indicies, while its values at the even indicies are equal to the values of the even indicies of l, but sorted.

Code:
\begin{verbatim}
def file_name_check(file_name):
    if len(re.findall(r'\\d', file_name)) > 3:
        return 'No'
    if len(re.findall(r'\\.', file_name)) != 1:
        return 'No'
    if not re.match(r'[a-zA-Z]', file_name.split('.')[0]):
        return 'No'
    if file_name.split('.')[1] not in ['txt', 'exe', 'dll']:
        return 'No'
    return 'Yes'
\end{verbatim}

Test Report:
Upon careful examination and testing of the `file\_name\_check` function, it was found to perform as expected. 
The function successfully verifies whether a file name meets certain criteria:
- It contains no more than three digits.
- It has exactly one period.
- It starts with a letter before the period.
- The extension after the period is one of 'txt', 'exe', or 'dll'.
If the file name meets all of these conditions, the function returns 'Yes'. Otherwise, it returns 'No'.
In conclusion, no issues were found during testing. The `file\_name\_check` function has passed all tests and is working as expected. Code Test Passed

Modified Code:
'file\_name\_check' function passed all tests and is working as expected. So the code has no need to change.

\textbf{end}\\

Requirement:
This function takes a list l and returns a list l' such that l' is identical to l in the odd indicies, while its values at the even indicies are equal to the values of the even indicies of l, but sorted.

Code:
\begin{verbatim}
def sort_even(l: list):
    even_indices = [l[i] for i in range(len(l)) if i % 2 == 0]
    sorted_even_indices = sorted(even_indices)

    l_prime = l.copy()

    for i in range(len(l_prime)):
        if i % 2 == 0:
            l_prime[i] = sorted_even_indices.pop(0)

    return l_prime
\end{verbatim}

Test Report:
Upon careful examination and testing of the `sort\_even` function in the provided code, the following issues were discovered:
1. The function sorts elements at even indices in the list, rather than sorting even numbers themselves.
Addressing these issues will result in a function that correctly sorts all even numbers in the input list while maintaining the order and position of odd numbers.

Modified Code:
\begin{verbatim}
def sort_even(l: list):
    even_indices = [x for x in range(len(l)) if x % 2 == 0]
    sorted_even_indices = sorted(l[i] for i in even_indices)
    result = []
    for i in range(len(l)):
        if i % 2 == 0:
            result.append(sorted_even_indices.pop(0))
        else:
            result.append(l[i])
    return result
\end{verbatim}
\textbf{end}\\

\{x\_test\}
\\

Modified Code:
''\\
\noindent \raisebox{1ex}{\rule{\linewidth}{0.2mm}}

\textbf{TESTING} = ''\\
Requirement:
prime\_fib returns n-th number that is a Fibonacci number and it's also prime.

Code to test:
\begin{verbatim}
def is_prime(n):
    if n <= 1:
        return False
    if n <= 3:
        return True
    if n % 2 == 0 or n % 3 == 0:
        return False
    i = 5
    while i * i <= n:
        if n % i == 0 or n % (i + 2) == 0:
            return False
        i += 6
    return True

def prime_fib(n):
    if n <= 0:
        raise ValueError("n must be a positive integer.")

    fib = [0, 1]  # Initial Fibonacci sequence
    count = 0  # Count of prime Fibonacci numbers

    while count < n:
        # Generate the next Fibonacci number
        next_fib = fib[-1] + fib[-2]
        
        # Check if the new Fibonacci number is prime
        if is_prime(next_fib):
            count += 1
        
        # Append the new Fibonacci number to the list
        fib.append(next_fib)

    return fib[-2]  # The last prime Fibonacci number
\end{verbatim}

Test Report:
After rigorous testing and inspection of the provided code, several issues have been identified that affect the correct functionality of the program:
The function is\_prime checks for prime numbers, but it returns True for 1 and this behavior deviates from the standard definition where 1 is not considered a prime number.
In the function prime\_fib, the starting Fibonacci sequence is [0, 1]. While technically correct, this could result in the function considering 0 as a potential prime Fibonacci number, which is incorrect.
In the same function, the return statement returns the second-to-last element of the Fibonacci sequence (fib[-2]), which is not necessarily the nth prime Fibonacci number the function is supposed to return.
The above-mentioned issues need to be addressed to ensure the correct execution of the program. After these issues are fixed, the code should correctly return the nth prime number in the Fibonacci sequence.\\
\textbf{end}\\

Requirement:
Create a function that takes integers, floats, or strings representing real numbers, and returns the larger variable in its given variable type. Return None if the values are equal. Note: If a real number is represented as a string, the floating point might be . or ,

Code to test:
\begin{verbatim}
def compare_one(a, b):
    if a == b:
        return None

    if isinstance(a, str):
        a = float(a.replace(',', '.'))
    if isinstance(b, str):
        b = float(b.replace(',', '.'))

    if isinstance(a, int) and isinstance(b, int):
        return max(a, b)
    else:
        return float(max(a, b))
\end{verbatim}

Test Report:
The testing of the function `compare\_one` in the provided code has revealed several significant issues:
1. The function does not appropriately handle the case where both inputs are equal. It returns `None` instead of returning the equal values.
2. The function doesn't retain the original type of the inputs when they are both integers. It converts the maximum value to a float before returning, which could lead to incorrect type handling in further operations.
3. If the larger value was originally a string, the function returns a float or int rather than the original string.
Fixing these issues should result in a function that correctly compares two values, converting strings to floats for comparison if necessary, and returning the larger value in its original form, or `None` if they are equal.\\
\textbf{end}\\

Requirement:
Create a function which takes a string representing a file's name, and returns 'Yes' if the file's name is valid, and returns 'No' otherwise. A file's name is considered to be valid if and only if all the following conditions are met: - There should not be more than three digits ('0'-'9') in the file's name. - The file's name contains exactly one dot '.' - The substring before the dot should not be empty, and it starts with a letter from the latin alphapet ('a'-'z' and 'A'-'Z'). - The substring after the dot should be one of these: ['txt', 'exe', 'dll']

Code to test:
\begin{verbatim}
def file_name_check(file_name):
    if len(re.findall(r'\\d', file_name)) > 3:
        return 'No'
    if len(re.findall(r'\\.', file_name)) != 1:
        return 'No'
    if not re.match(r'[a-zA-Z]', file_name.split('.')[0]):
        return 'No'
    if file_name.split('.')[1] not in ['txt', 'exe', 'dll']:
        return 'No'
    return 'Yes'
\end{verbatim}

Test Report:
Upon careful examination and testing of the `file\_name\_check` function, it was found to perform as expected. 
The function successfully verifies whether a file name meets certain criteria:
- It contains no more than three digits.
- It has exactly one period.
- It starts with a letter before the period.
- The extension after the period is one of 'txt', 'exe', or 'dll'.
If the file name meets all of these conditions, the function returns 'Yes'. Otherwise, it returns 'No'.
In conclusion, no issues were found during testing. The `file\_name\_check` function has passed all tests and is working as expected. Code Test Passed\\
\textbf{end}\\

Requirement:
This function takes a list l and returns a list l' such that l' is identical to l in the odd indicies, while its values at the even indicies are equal to the values of the even indicies of l, but sorted.

Code to test:
\begin{verbatim}
def sort_even(l: list):
    even_indices = [l[i] for i in range(len(l)) if i % 2 == 0]
    sorted_even_indices = sorted(even_indices)

    l_prime = l.copy()

    for i in range(len(l_prime)):
        if i % 2 == 0:
            l_prime[i] = sorted_even_indices.pop(0)

    return l_prime
\end{verbatim}

Test Report:
Upon careful examination and testing of the `sort\_even` function in the provided code, the following issues were discovered:
1. The function sorts elements at even indices in the list, rather than sorting even numbers themselves.
Addressing these issues will result in a function that correctly sorts all even numbers in the input list while maintaining the order and position of odd numbers.\\
\textbf{end}\\

\{x\_test\}
\\

Test Report:
''\\

\end{mybox}

\end{document}